\documentclass[%
 reprint,preprintnumbers,
nofootinbib,
 amsmath,amssymb,
 aps
prd,
]{revtex4-1}
\pdfoutput=1

\usepackage{graphicx}
\usepackage[utf8]{inputenc}
\usepackage{flushend}
\usepackage{dcolumn}
\usepackage{bm}
\usepackage{balance}

\usepackage[normalem]{ulem}
\usepackage[colorlinks = true,
            linkcolor = blue,
            urlcolor  = blue,
            citecolor = blue,
            anchorcolor = blue]{hyperref}
\usepackage{verbatim}
\usepackage{color,ulem}
\usepackage[english]{babel}
\usepackage{MnSymbol,wasysym}
\usepackage[utf8]{inputenc}
\input Starburst.fd
\newcommand*\initfamily{\usefont{U}{Starburst}{xl}{n}}\initfamily 

\newcommand{\beq}{\begin{eqnarray}}
\newcommand{\eeq}{\end{eqnarray}}
\usepackage{amsmath}
\usepackage{tikz}
\usetikzlibrary{decorations.pathmorphing}
\usetikzlibrary{shapes.misc}
\tikzset{cross/.style={cross out, draw=black, minimum size=8*(#1-\pgflinewidth), inner sep=0pt, outer sep=0pt},
cross/.default={1pt}}
\usetikzlibrary{patterns,math}
\definecolor{applegreen}{rgb}{0.55, 0.71, 0.0}

\newcommand{\td}{\text{d}}

\renewcommand{\d}{\delta}

\newcommand{\dd}{\mathrm{d}}

\newcommand{\ke}[1]{{ #1}\rangle}

\def\kd#1{\left[#1\right]}
\def\ke#1{\left\{#1\right\}}

\begin{document}
\preprint{IFT-UAM/CSIC-23-54}

\title{Entanglement entropy analysis of dyonic black holes using doubly holographic theory}

\author{Hyun-Sik Jeong$^{1,2}$}\email{hyunsik.jeong@csic.es}
\author{Keun-Young Kim$^{3,4}$}\email{fortoe@gist.ac.kr}
\author{Ya-Wen Sun$^{5,6}$}\email{yawen.sun@ucas.ac.cn}

\affiliation{$^{1}$Instituto de F\'isica Te\'orica UAM/CSIC, Calle Nicol\'as Cabrera 13-15, 28049 Madrid, Spain}
\affiliation{$^{2}$Departamento de F\'isica Te\'orica, Universidad Aut{\'o}noma de Madrid, Campus de Cantoblanco, 28049 Madrid, Spain}
\affiliation{$^{3}$Department of Physics and Photon Science, Gwangju Institute of Science and Technology,
123 Cheomdan-gwagiro, Gwangju 61005, Korea}
\affiliation{$^{4}$Research Center for Photon Science Technology, Gwangju Institute of Science and Technology, 123 Cheomdan-gwagiro, Gwangju 61005, Korea}
\affiliation{$^{5}$School of Physical Sciences, University of Chinese Academy of Sciences, Zhongguancun east road 80, Beijing 100190, China}
\affiliation{$^{6}$Kavli Institute for Theoretical Sciences, University of Chinese Academy of Sciences, Zhongguancun east road 80, Beijing 100049, China}

\begin{abstract}
We investigate the entanglement between the eternal black hole and Hawking radiation. For this purpose, we utilize the doubly holographic theories and study the \textit{entanglement entropy} of the radiation to find the Page curve consistent with the unitarity principle. Doubly holographic theories introduce two types of boundaries in the AdS bulk, namely the usual AdS boundary and the Planck brane. In such a setup, we calculate the entanglement entropy by examining two extremal surfaces: the Hartman-Maldacena (HM) surface and the island surface. The latter surface emerges when the island appears on the Planck brane.
In this paper, we provide a detailed analysis of dyonic black holes with regard to the Page curve in the context of the doubly holographic setup. 
To begin with, we ascertain that the pertinent topological terms must be included in the Planck brane to describe the systems at finite density and magnetic field.
Furthermore, we also develop a general numerical method to compute the time-dependent HM surface and achieve excellent agreement between the numerical results and analytical expressions.
Utilizing numerical methodology, we find that the entanglement entropy of dyonic black holes exhibits unitary evolution over time, wherein it grows in early time and reaches saturation after the Page time. The initial growth can be explained by the HM surface, while the saturation is attributed to the island surface.
In addition, using the holographic entanglement density, we also show that, for the first time, the saturated value of the entanglement entropy is twice the Bekenstein-Hawking entropy with the tensionless brane in double holography.
\end{abstract}

\maketitle
%
\section{Introduction}

The black hole information paradox has been a long-standing and one of the central problems in theoretical physics~\cite{Hawking:1975vcx,Hawking:1976ra,Almheiri:2020cfm,Raju:2020smc,Harlow:2014yka}. An important aspect of this issue is to understand the entanglement between the black hole and the radiation in a unitary fashion. In order to demonstrate that the black hole plus radiation system behaves as a unitary quantum system, one may need to show that the time evolution of entanglement entropy of the radiation follows a characteristic feature of unitary quantum systems, the Page curve~\cite{Page:1993wv,Page:2013dx}.

In the case of the \textit{evaporating} black holes, the Page curve coherent with the unitarity principle exhibits that the von Neumann (or fine-grained) entropy gives the initial rise of the Page curve due to the early Hawking radiation, and then decreases after the Page time.
Recall that Hawking's earlier calculation~\cite{Hawking:1974rv} was stating that the entropy keeps growing until the black holes entirely evaporate.

On the other hand, for the case of the \textit{eternal} black holes, the unitarity requires that the growth of the entropy stops at the Page time and the Page curve is upper bounded by $2 S_{\text{BH}}$, {where} $S_{\text{BH}}$ is the Bekenstein-Hawking entropy of the black hole, since the fine-grained entanglement entropy cannot exceed the coarse-grained black hole entropy~\cite{Bekenstein:1980jp,Almheiri:2020cfm}.

From the perspective of quantum mechanics, the unitarity-inherited theory, it is clear that the expected Page curve can happen. Nevertheless, one would also like to understand how the Page curve can be achieved from the gravity point of view. 
In recent years, the holographic principle (or the AdS/CFT correspondence) provides a great breakthrough to understand the Page curve along this direction through the study of the emergence of the islands and quantum extremal surfaces~\cite{Penington:2019npb,Almheiri:2019hni,Almheiri:2019psf,Almheiri:2019psy,Penington:2019kki}.
The inclusion of new bulk regions known as ``islands" 
after the Page time played a significant role in reproducing the Page curve, i.e., bending/saturating the growing entanglement entropy for the evaporating/eternal black holes.
In particular, motivated by the Ryu-Takayanagi (RT) formula together with its generalization~\cite{Ryu:2006bv,Ryu:2006ef,Hubeny:2007xt,Lewkowycz:2013nqa}, the gravitational analysis has been facilitated by the development of the holographic computations of the fine-grained entropy of a system by the quantum extremal surface~\cite{Engelhardt:2014gca}.
\begin{figure*}[]
\centering
     {\includegraphics[width=15.0cm]{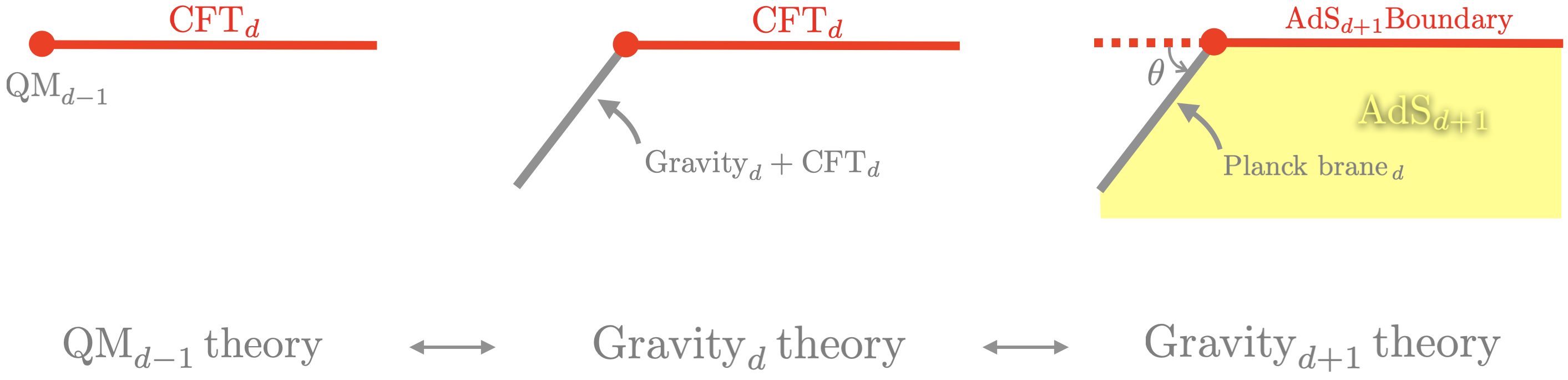} \label{}}
 \caption{A sketch of three different representations of the same system where $d$ denotes the dimension. \textbf{The \textit{left}} is the quantum mechanical (QM) description where QM lives at the boundary of the bath CFT (red). \textbf{The \textit{center}} is the gravity theory plus the matter CFT (gray), coupled to a bath having the same CFT (red): the CFT is assumed to have a holographic dual. \textbf{The \textit{right}} is a one higher-dimensional holographic description of \textbf{the \textit{center}} in which the d-dimensional matter CFT (gray) is replaced by (d+1) dimensional ambient AdS spacetime (yellow region). Note that the d-dimensional gravity is described by the Planck brane (i.e., a dynamical boundary metric on the brane) in the AdS spacetime and the bath CFT corresponds to the boundary of the AdS spacetime. In short, \textbf{the \textit{right}} consists of two boundaries (Planck brane, AdS boundary) in the AdS spacetime.}\label{SK1}
\end{figure*}

The essential idea behind gravity computations is that the Hawking radiation is absorbed by a non-gravitational bath coupled to the asymptotic boundary of the gravitational system containing the black hole. For instance, the black hole in AdS spacetime is connected to a flat space on the boundary, which is treated as a thermal bath in order to collect the radiation.
Then, one can determine the entanglement entropy of the radiation, $S_{\text{R}}$, by the ``island formula" as
\begin{align}\label{ISFO1}
\begin{split}
S_{\text{R}}=\min_{\text{I}} \ke{ \mathop{\text{ext}}\limits_{\text{I}} \kd{ S[\text{R} \cup \text{I}]+\frac{\operatorname{Area}[\partial \text{I}]}{4 G_{N}}}},
\end{split}
\end{align}
where $G_N$ is the gravitational constant. Note that \eqref{ISFO1} takes into account the entanglement entropy of the radiation region ($\text{R}$) together with the gravitational bulk region called Islands ($\text{I}$).
Also note that $S_{\text{R}}$ is determined by the standard procedure, i.e., when the entire function gets minimized after taking the extremization of all possible islands.
For instance, for the case of evaporating black holes, 
the entropy \eqref{ISFO1} at the early time is evaluated without the inclusion of any islands, and the result agrees with Hawking's calculation. 
However, the contribution of islands becomes more prominent over time, leading to the appearance of a new saddle point during the minimization of \eqref{ISFO1} in the later time.\footnote{This phenomenon arises from the fact that the quanta of Hawking radiation possess a significant degree of entanglement with the quantum fields located beyond the black hole horizon.}
At this stage, the black hole entropy, which appears in the second term of \eqref{ISFO1}, dominates the entropy computation and produces the expected Page curve. 
Utilizing the island formula, the Page curve has been extensively developed and investigated under various scenarios, for instance~\cite{Engelhardt:2014gca,Almheiri:2019hni,Almheiri:2019psf,Almheiri:2019psy,Almheiri:2019yqk,Ling:2020laa,Penington:2019npb,Sully:2020pza,Chen:2019iro,Anegawa:2020ezn,Balasubramanian:2020hfs,Gautason:2020tmk,Hartman:2020swn,Hollowood:2020cou,Alishahiha:2020qza,Rozali:2019day,Hashimoto:2020cas,Karananas:2020fwx,Wang:2021woy,Kim:2021gzd,Ahn:2021chg,Yu:2021cgi,Geng:2020qvw,Bak:2020enw,Verheijden:2021yrb,Li:2020ceg,Chandrasekaran:2020qtn,Hollowood:2020kvk,Bousso:2019ykv,Akers:2019nfi,Liu:2020gnp,Bousso:2020kmy,Chen:2020jvn,Hartman:2020khs,Sybesma:2020fxg,Balasubramanian:2020xqf,Chen:2019uhq,Chen:2020uac,Chen:2020hmv,Hernandez:2020nem,He:2021mst,Grimaldi:2022suv,KumarBasak:2020ams,Kawabata:2021hac,Matsuo:2020ypv,Krishnan:2020fer,Caceres:2020jcn,Geng:2021wcq,Anderson:2020vwi,Bhattacharya:2021jrn,Yu:2021rfg,Ageev:2022qxv,Omidi:2021opl,Tian:2022pso,Lu:2022tmt,HosseiniMansoori:2022hok,Miao:2023unv,Luongo:2023jyz,RoyChowdhury:2023eol,Ageev:2023mzu}.\footnote{The provided list is not exhaustive. We encourage readers to refer to the references in aforementioned literature to explore the related topic further.}\\

\noindent \textbf{The doubly holographic theories and Page curve.}
In particular, in the context of \textit{doubly holographic theories} (which are closely related to AdS/BCFT~\cite{Takayanagi:2011zk,Fujita:2011fp,Nozaki:2012qd,Miao:2018qkc,Miao:2017gyt,Chu:2017aab,Chu:2021mvq} and brane world theory~\cite{Randall:1999ee,Randall:1999vf,Karch:2000ct}), a useful method has been developed in \cite{Almheiri:2019hni} for holographic computation of the entanglement entropy of Hawking radiation.\footnote{See also \cite{Almheiri:2019hni,Almheiri:2019yqk,Rozali:2019day,Chen:2019uhq,Almheiri:2019psy,Kusuki:2019hcg,Chen:2019iro,Balasubramanian:2020hfs,Gautason:2020tmk,Alishahiha:2020qza,Geng:2020qvw,Chen:2020uac,Bousso:2020kmy,Krishnan:2020fer,Chen:2020jvn,Ling:2020laa, Hernandez:2020nem,Geng:2020fxl,Kawabata:2021hac,Kawabata:2021vyo,Geng:2021hlu,Akal:2020wfl,Miao:2020oey,Sully:2020pza,Neuenfeld:2021bsb,Chen:2020hmv,Ghosh:2021axl,Omiya:2021olc,Bhattacharya:2021nqj,Geng:2021mic,Geng:2022tfc,Sun:2021dfl,Chou:2021boq,Wang:2021xih,Caceres:2020jcn,Geng:2021iyq,Ling:2021vxe,Liu:2022ezb,Hu:2022lxl,Grimaldi:2022suv,Anous:2022wqh, Basu:2022reu,Liu:2022pan,Lee:2022efh,Uhlemann:2021nhu,Karch:2022rvr,Yadav:2022mnv,Hu:2022ymx,Hu:2022zgy,Miao:2022mdx,Miao:2023unv,Perez-Pardavila:2023rdz,Karch:2023ekf,Bhattacharya:2023drv,Afrasiar:2022ebi,Afrasiar:2022fid,Afrasiar:2023jrj,Azarnia:2021uch,RoyChowdhury:2022awr,Geng:2019bnn,Geng:2023iqd,Geng:2023qwm} and the references therein for the recent development of various quantum information quantities (such as entanglement entropy, reflected entropy, complexity, and negativity etc), resorting to the doubly holographic theories.}
Within the doubly holographic framework (i.e., the gravity $+$ matter theory where the matter sector has one higher-dimensional holographic dual; see also the sketch in Fig. \ref{SK1}), the authors in \cite{Almheiri:2019hni} showed that the prescription for extremizing the generalized entropy \eqref{ISFO1} can be equivalent to the standard RT/HRT prescription~\cite{Ryu:2006bv,Hubeny:2007xt} of extremizing the area. In other words, following a Randall-Sundrum type with a d-dimensional brane in a (d+1) dimensional ambient spacetime~\cite{Randall:1999vf,Karch:2000ct,Dvali:2000hr}, the quantum extremal surfaces in d-dimension corresponds to the standard RT surfaces in (d+1) dimensions. Thus, it becomes a feasible gravity calculation for the entanglement entropy. We will review this procedure in detail in the next section.

Furthermore, considering evaporating black holes, the authors in \cite{Almheiri:2019hni} ensured that the minimal surface of the Hawking radiation coincides with that of the evaporating black holes (so that we can focus on evaluating the entanglement entropy of the radiation in order to investigate the entanglement between the radiation and black hole).\footnote{From the recent development beyond the scope of doubly holographic theories~\cite{Ageev:2022qxv,Ageev:2023mzu}, one should exercise caution with regard to the complementarity property, whereby the entropy of the radiation may not be equivalent to that of its complement. We would like to thank {D.S. Ageev, I.Ya. Aref'eva, A.I. Belokon, V.V. Pushkarev, T.A. Rusalev} for the related comments.} Recall that when the combined state of the black hole and Hawking radiation is a pure state, the entanglement entropy of the radiation should be equivalent to that of the black hole.\footnote{The essence of this doubly holographic approach \cite{Almheiri:2019hni} is that the interior region of the black hole may be connected to the radiation through the additional dimension. The entanglement between the interior modes of the quantum matter and the Hawking radiation can be linked via the geometric connection. In this regard, the extra dimension can be seen as a demonstration of the ER$=$EPR~\cite{Maldacena:2013xja} concept. See also \cite{Susskind:2012uw,Papadodimas:2012aq}.}
\begin{figure*}[]
\centering
     {\includegraphics[width=13.0cm]{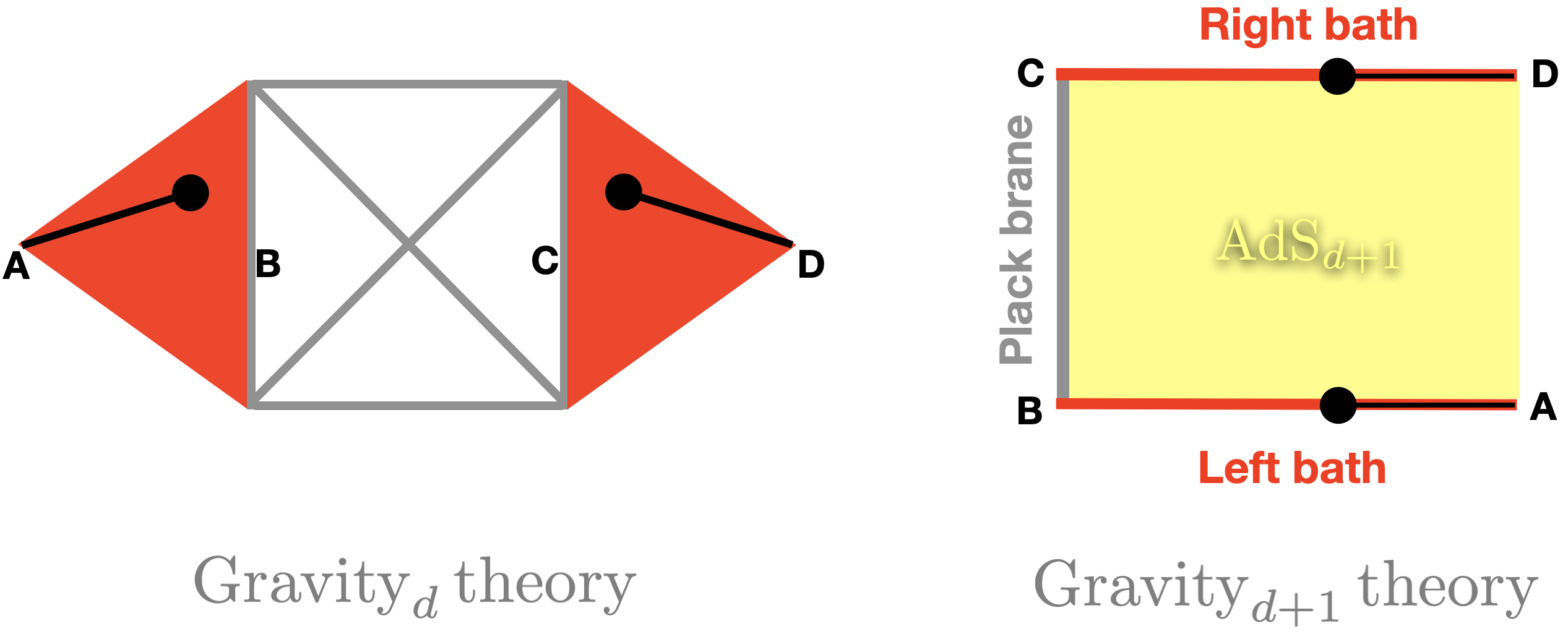} \label{}}
 \caption{A sketch of the two-sided eternal black holes: (A, B, C, D) points may be useful for the readers to associate \textbf{the \textit{left}} and \textbf{the \textit{right}}. \textbf{The \textit{left}} is composed of the d-dimensional black hole with the conformal matter living inside (gray). The black lines represent the radiation regions in the two left/right baths (red). \textbf{The \textit{right}} is a one higher-dimensional description where the d-dimensional black hole is described by the Planck brane and the conformal matter is replaced by the AdS spacetime (yellow region). Here the angle $\theta$ is taken to be $\pi/2$. We will elaborate on this point in the next section.}\label{SK2}
\end{figure*}

It is worth noticing that \cite{Almheiri:2019hni} facilitated the entanglement entropy calculation of the evaporating two-dimensional black holes using the doubly holographic setup, i.e., $d=2$ in Fig. \ref{SK1}, and left the analysis of the higher-dimensional black holes as future investigation. For this purpose, \cite{Almheiri:2019psy} initiated the study of the higher-dimensional case for the case of eternal black holes in five-dimensional Schwarzschild-AdS black holes within the doubly holographic setup. Note that for the two-sided eternal black holes, Fig. \ref{SK1} can be expressed as Fig. \ref{SK2}.
Furthermore, the case of the Reissner-Nordström-AdS black hole is investigated in the four-dimensional eternal black holes in \cite{Ling:2020laa}.
The upshot of this analysis with the higher-dimensional black hole is that using the ordinary RT/HRT prescription, the doubly holographic theories can provide the affirmative result for the resolution of the information paradox by virtue of the emergence of an island. In addition, one can also produce the Page curve consistent with the unitarity principle even for the higher-dimensional (neutral or charged) black holes.\\

\noindent \textbf{Motivation of this paper.}
In this paper, we further investigate the information paradox in the higher-dimensional eternal black holes using the doubly holographic theories, i.e., we intend to show that the island paradigm would be a general solution to the information paradox for black holes in higher dimensions.

In particular, we consider \textit{dyonic} Reissner-Nordström-AdS black hole in the same dimension of \cite{Ling:2020laa}.
There are several motivations to consider this dyonic black hole in the doubly holographic setup. Most importantly, studying the magneto-transport properties of the dyonic black holes, the authors in \cite{Fujita:2012fp,Melnikov:2012tb} claimed that a finite charge density \textit{must} be supported by a magnetic field within AdS/BCFT construction (the same gravity setup in doubly holography).

In other words, this implies that one may not consider the finite density system in the framework of the doubly holographic theories without introducing an external magnetic field. Apparently, this seems to be in contrast with \cite{Ling:2020laa} since the finite charge density effect is investigated there even without a magnetic field.
Therefore, the scope of this work not only extends the analysis in \cite{Ling:2020laa} to include the external magnetic fields but also aims to reconcile this would-be disagreement.

Furthermore, giving all the details of the computations, we also study how the doubly holographic theories can produce the double Bekenstein-Hawking entropy, $2 S_{\text{BH}}$, at late times in the Page curve. Notice that although \cite{Almheiri:2019psy,Ling:2020laa} showed that the entanglement entropy of the eternal black holes is saturated at late times, its value was not comparable with $2 S_{\text{BH}}$ as it is supposed to be for the eternal black holes. In this paper, considering the entanglement density concept~\cite{Gushterov:2017vnr,Erdmenger:2017pfh,Giataganas:2021jbj,Jeong:2022zea}, we provide the possible way to obtain $2 S_{\text{BH}}$ within the doubly holographic setup.\\

This paper is organized as follows. 
In section \ref{sec2la}, we review the doubly holographic setup for dyonic black holes. 
In section \ref{sec3la}, we present the formula of the entanglement entropy of the Hawking radiation in the framework of doubly holographic theories introduced in section \ref{sec2la}.
In section \ref{sec4la}, we study the extremal surfaces of dyonic black holes and discuss the Page curve. In addition, considering the entanglement density, we provide a way to exhibit the double Bekenstein-Hawking entropy within doubly holographic theories.
Section \ref{sec5la} is devoted to conclusions.

%
\section{The doubly holographic setup: a quick review}\label{sec2la}
In this section, following \cite{Almheiri:2019hni} we introduce the doubly holographic setup for dyonic black holes: $d=3$ in Fig. \ref{SK2}.
In other words, we consider 3-dimensional (electrically/magnetically) charged eternal black holes coupled to two baths on each side where the conformal matter lives in the bulk: \textbf{The \textit{left}} configuration in Fig. \ref{SK2}.

As demonstrated in the introduction, this configuration can be equivalently described by a doubly holographic setup, i.e., a 3-dimensional black hole is replaced by the Planck brane and the conformal matter is dual to a 4-dimensional AdS spacetime: \textbf{The \textit{right}} configuration in Fig. \ref{SK2}.
Thus, in the doubly-holographic setup, we are led to consider the action of the dyonic black holes~\cite{Fujita:2012fp,Melnikov:2012tb}, $S_{total}$, as 
\begin{align}\label{TOTAC}
\begin{split}
S_{total} &= S_{bulk} + S_{brane} \,, \\
S_{bulk} &= \frac{1}{16 \pi G_N}\int \dd^4 x\, \sqrt{-g}\left(R+\frac{6}{L^2}\right) \\
 &\qquad - \,\frac{1}{4}\int \dd^4 x\,\sqrt{-g}F_{\mu\nu}F^{\mu\nu}-\frac{\Theta}{8\pi^2}\int F\wedge F \,, \\
S_{brane} &= \frac{1}{8 \pi G_N}\int \dd^3 x\,\sqrt{-h}\left(K - \alpha \right) - \frac{k}{4\pi}\int A\wedge F \,,
\end{split}
\end{align}
where $G_N$ is the gravitational constant and $L$ the AdS radius.

The bulk action $S_{bulk}$ is composed of the metric $g_{\mu \nu}$ together with the gauge field $A_{\mu}$ via its field strength $F=\dd A${: see \eqref{BGMET}}. The last term in the bulk action, $\approx \Theta\int F\wedge F$, is a topological term (so it does {not} appear in the equations of motion) which is relevant for the analysis of the boundary conditions.

The other action, $S_{brane}$, is the one for the Planck brane where its induced metric or the extrinsic curvature on the Planck brane is denoted as $h_{ab}$ and $K$, respectively. Here, $\alpha$ is related to the tension on the Planck brane as we will show shortly. The last term in the brane action, $\approx k\int A\wedge F$, is a Chern-Simons term on the brane, which is suitable for the analysis of the dyonic black holes within AdS/BCFT or the doubly-holographic setup: see \cite{Fujita:2012fp,Melnikov:2012tb} for a more detailed description of it.\footnote{One can also introduce the supplemental two kinds of boundary actions into the total action \eqref{TOTAC}. The first one would be the usual Gibbons-Hawking term {on the AdS boundary} and the other is the junction term at the intersection of the Planck brane and the AdS boundary (i.e., at the red point in Fig. \ref{ST3}). 
In this paper, we omit these additional terms to avoid clutter, which would be superficial for our discussion. For the readers who are interested, please refer to \cite{Fujita:2012fp,Melnikov:2012tb}.}

It is worth noticing that the topological terms ($\Theta$ and $k$) were not taken into account for the analysis of the electrically charged black holes in \cite{Ling:2020laa}. As we will show, these topological terms may play an important role to investigate the aspect of the doubly holographic theories even in the case of electrically charged black holes.

\subsection{The Planck brane and Neumann boundary conditions}
The bulk equation of motion from the action \eqref{TOTAC} reads
\begin{align}\label{BGEOM1}
\begin{split}
 R_{\mu\nu} - \frac{R}{2}\,g_{\mu\nu} - \frac{3}{L^2}\,g_{\mu\nu} &= 8 \pi G_N  \left(F_{\mu\rho}{F_{\nu}}^{\rho} - \frac{g_{\mu\nu}}{4} F_{\rho\sigma}F^{\rho\sigma} \right) \,, \\
\nabla^\mu F_{\mu\nu} &= 0 \,.
\end{split}
\end{align}
Furthermore, in addition to the bulk equations of motion above, it is also required to specify the boundary conditions in order to establish a well-defined variational principle in a space with boundaries: recall that in the doubly holographic theories, there can be two kinds of boundaries (the AdS boundary and the Planck brane). 

Following the standard holographic duality, Dirichlet boundary conditions are imposed on the AdS boundary.\footnote{See \cite{Ahn:2022azl,Jeong:2023las,Ishibashi:2023luz,Baggioli:2023oxa,Harada:2023cfl} and references therein for the recent development of the mixed boundary conditions on the AdS boundary.} 
On the other hand, in AdS/BCFT or doubly holographic setup, Neumann boundary conditions are imposed on the Planck brane.
To discuss the boundary condition on the Planck brane, one can find the following boundary terms by a variation of the total action with respect to the metric/gauge field, respectively as
\begin{align}\label{BDRYT}
\begin{split}
 \frac{1}{16 \pi G_N} & \int_{pb} \d^3 x\, \sqrt{-h}\Big[K_{ab}-(K-\alpha)h_{ab}\Big]\delta h^{ab} \,, \\
2&\int_{pb} \Bigg[\frac12\,\ast F - \left(\frac{\Theta}{8\pi^2}+\frac{k}{4\pi}\right)F\Bigg] \wedge \delta A \,,
\end{split}
\end{align}
where $pb$ indicates that these are the objects on the Planck brane.

From the equations in \eqref{BDRYT}, one can impose Dirichlet boundary conditions (i.e., $\delta h^{ab}=\delta A=0$). However, one may need to employ Neumann boundary conditions to determine the Planck brane dynamically, i.e.,
\begin{align}\label{BDRYT2}
\begin{split}
K_{ab}-(K-\alpha)h_{ab} = 0 \,, \quad \frac12\,\ast F -\left(\frac{\Theta}{8\pi^2}+\frac{k}{4\pi}\right)F  = 0 \,.
\end{split}
\end{align}
Note that imposing the Neumann boundary condition (rather than the Dirichlet one) allows a specific boundary component of the bulk to be referred to as a Planck brane or RS brane~\cite{Randall:1999ee,Randall:1999vf}.\footnote{The string theory orientifold construction may also support this Neumann boundary condition as a natural choice for the boundary condition~\cite{Fujita:2011fp}.}\\

\begin{figure}[]
\centering
     {\includegraphics[width=8.0cm]{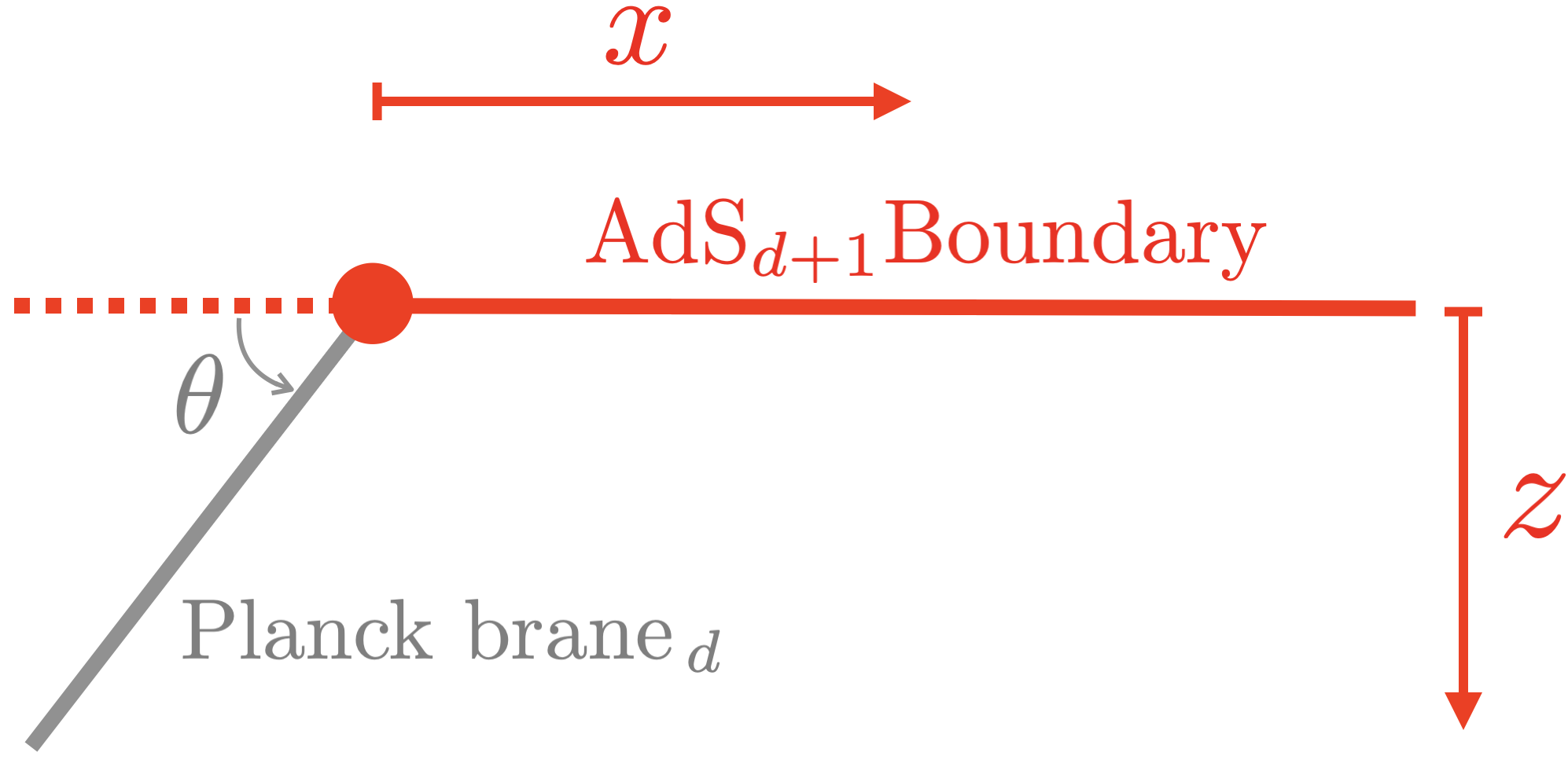} \label{}}
 \caption{A simple setup of a Planck brane or Randall-Sundrum brane. The Planck brane (gray line) is anchored at $(z, x) = (0, 0)$ and penetrates into the bulk with an angle $\theta$.}\label{ST3}
\end{figure}

Next, let us review the implication of the Neumann boundary conditions \eqref{BDRYT2}.
When the bulk geometry is asymptotically AdS such as 
\begin{equation}\label{}
\dd s^2 \approx \frac{L^{2}}{z^2} \left[- \dd t^2 +  \dd z^2 +  \dd x^2 +  \dd y^2  \right] \,,
\end{equation}
where the AdS boundary is located at $z=0$, the Planck brane can be described by the hypersurface
\begin{align}\label{PLOC}
\begin{split}
z + x \tan \theta = 0 \,.
\end{split}
\end{align}
Here $\theta$ is the angle between the AdS boundary and the Planck brane. See Fig. \ref{ST3}.
Then, one can evaluate the extrinsic curvature on this hypersurface as 
\begin{align}\label{}
\begin{split}
K_{ab} = \frac{\cos \theta}{L} h_{ab} \,.
\end{split}
\end{align}
Plugging it into the first boundary condition in \eqref{BDRYT2}, one finds that the parameter $\alpha$ is determined by the angle $\theta$ 
\begin{align}\label{TAR}
\begin{split}
\alpha = \frac{2 \cos \theta}{L} \,.
\end{split}
\end{align}
Note that from the Israel junction condition~\cite{Israel:1966rt}, such a quantity, $\alpha$, can be interpreted as the tension of the brane.
In other words, the first Neumann condition in \eqref{BDRYT2} produces \eqref{TAR} implying that the angle $\theta$ sets the tension of the Planck brane in which $\theta=\pi/2$ gives the tensionless brane.

One can also find the same result with a unit vector normal,  $n^\mu$, to the Planck brane. See the explicit form of $n^{\mu}$ in \cite{Fujita:2012fp,Melnikov:2012tb}.
Using the defined $n^\mu$ together with the pullback of the equations \eqref{BDRYT2} to the bulk, one can find that the first Neumann condition in \eqref{BDRYT2} produces  
\begin{align}\label{}
\begin{split}
 x'(z)  = \frac{L \alpha}{\sqrt{4 - L^2 \alpha^2}} \,.
\end{split}
\end{align}
Then, plugging \eqref{PLOC} into this equation, we find the same result with \eqref{TAR}.

Similar to the discussion with the Neumann condition from the variation of the metric above, one can also study what the Neumann condition from the variation of the gauge field implies, i.e., the second Neumann condition in \eqref{BDRYT2}.

For this purpose, we consider the gauge field as in \eqref{BGMET} where $A_t = \mu - \rho z$.
%
%
Here $\mu$ is the chemical potential, $\rho$ a density, and $B$ an external magnetic field.
Then, using $n^{\mu}$ again, the second Neumann condition in \eqref{BDRYT2} can be rewritten as
\begin{align}\label{SNEQ}
\begin{split}
\sqrt{g}\,n_\nu F^{\nu\mu} + \frac{c_{pb}}{2}\,n_\nu\epsilon^{\nu\mu\rho\sigma} F_{\rho\sigma}=0 \,, \quad c_{pb}:= \frac{\Theta}{4\pi^2}+\frac{k}{2\pi} \,,
\end{split}
\end{align}
where we define the coefficients of the topological terms on the Planck brane as $c_{pb}$.
Furthermore, given $n^{\mu}$ in \cite{Fujita:2012fp,Melnikov:2012tb}, one can find that \eqref{SNEQ} gives two equations as
\begin{align}\label{COND1}
\begin{split}
0  =  \rho\cos\theta +  c_{pb} B\cos\theta\,, \quad 0  =  B\sin\theta -  c_{pb} \rho\sin\theta\,.
\end{split}
\end{align}

Before proceeding, two remarks are in order. 
First, let us revisit the case of the purely electrically charged black hole~\cite{Ling:2020laa}.
Implementing \eqref{COND1} in the absence of the magnetic field, one is led to consider 
\begin{align}\label{FDR}
\begin{split}
(\rho\neq0\,, B=0): \quad \theta = \frac{\pi}{2} \,\,\,\,\&\,\,\,\, c_{pb} =0 \,,  
\end{split}
\end{align}
in order to study a finite density system. 
In other words, both the tension of the brane \eqref{TAR} and topological terms should vanish.\footnote{The analysis at the finite tension may be regarded by adding extra terms on the brane such as a Dvali-Gabadadze-Porrati term (DGP). However, the entanglement entropy at finite tension has been only approached at $t=0$ using the DeTurck method. Thus further future investigation or development for the time evolution of the entanglement entropy at finite tension is still required \cite{Almheiri:2019psy,Ling:2020laa}.}
This implies that in the doubly holographic framework, the purely electrically charge black holes may be investigated only at zero tension, rather than at a weak (but finite) tension as in \cite{Ling:2020laa}.\footnote{\label{ft9}The authors in \cite{Ling:2020laa} also considered \eqref{SNEQ} with $c_{pb}=0$. However, the role of the Neumann condition of the gauge field is not explored there (e.g., \eqref{FDR}) in detail. Note that if one tries to consider the finite tension ($\theta \neq \pi/2$) at $B=0$ as in \cite{Ling:2020laa}, \eqref{COND1} indicates that the system should be neutral $\rho=0$.}

Second, in this paper we consider the tensionless ``limit" (but a finite $c_{pb}$) to study the dyonic black holes as in \cite{Fujita:2012fp,Melnikov:2012tb} in order for the continuity from a finite density result \eqref{FDR}. When \eqref{COND1} at $\theta=\pi/2$, we have 
\begin{align}\label{RATD2}
\begin{split}
(\rho\neq0\,, B\neq0): \quad \theta = \frac{\pi}{2} \,\,\,\,\&\,\,\,\, \frac{\rho}{B} = \frac{1}{c_{pb}} \,,
\end{split}
\end{align}
which is one of the main features of the doubly holographic or the AdS/BCFT setup of dyonic black holes: the density $\rho$ and the magnetic field $B$ are no longer independent parameters by virtue of the additional boundary conditions at the Planck brane.\footnote{Based on the fact that the ratio $\rho/B$ correspond to the Hall conductivity of the dyonic black holes, \cite{Fujita:2012fp,Melnikov:2012tb} argued that the AdS/BCFT construction may provide the relevant holographic description similar to the Chern-Simon description of the quantum Hall effect since the Hall conductivity is independent of both $\rho$ and $B$, but inversely proportional to topological coefficients. See also \cite{Santos:2023flb} for a similar discussion in the presence of Horndeski gravity term.}
Also note that the tensionless Planck brane in doubly holography indicates that the brane can be considered a probe so that its backreaction to the background geometry is neglected.
\begin{figure*}[]
\centering
     {\includegraphics[width=14.0cm]{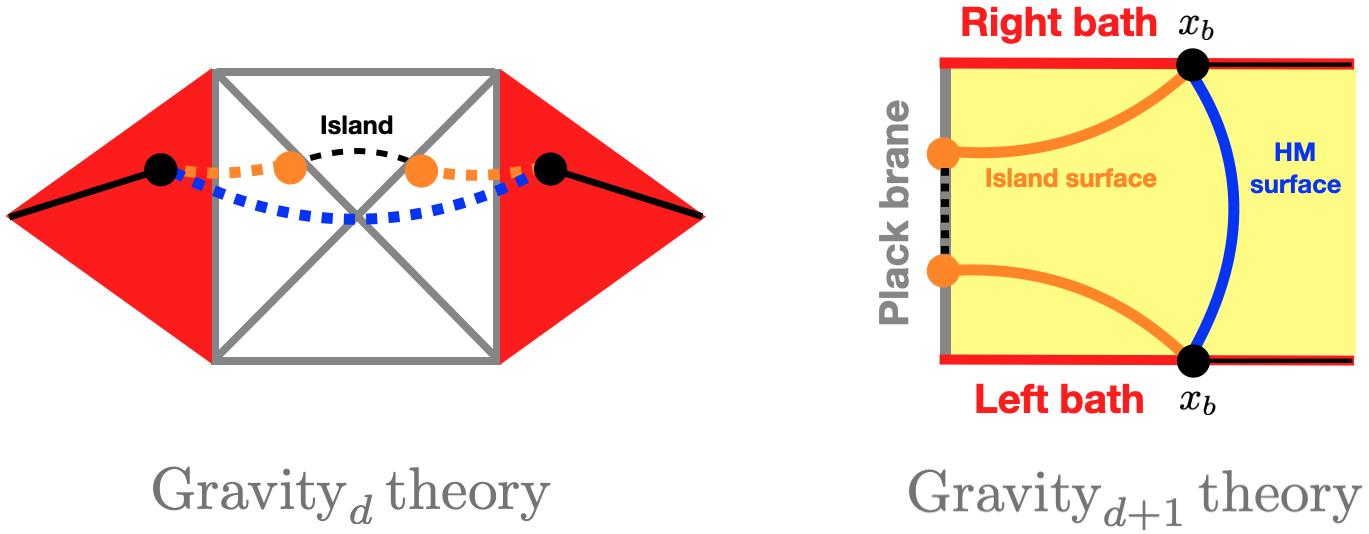} \label{}}
 \caption{A sketch of the two-sided eternal black holes with the radiation region (solid black lines) in the two baths (red regions). 
In \textbf{the \textit{left}}, the two candidates of the quantum extremal surfaces are expressed {in different colors}: the connected one (blue), the disconnected one (orange). The island is also depicted as the dashed black line.
In \textbf{the \textit{right}}, i.e., in the doubly holographic setup, the quantum extremal surfaces are {measured by} the HM surface (blue) or the island surface (orange). The endpoint of the radiation region in the bath, $x_b$, is represented by black dots.}\label{SK223FIG}
\end{figure*}

In the next section, using \eqref{RATD2} we will examine if the Page curve of the entanglement entropy in the doubly holographic theories can be produced even at finite $c_{pb}$. Note that it may not be straightforward to expect the effect of the topological coefficients on the Page curve without explicit computations. For instance, is the Page time suppressed or enhanced by a finite $c_{pb}$? When the Page time is suppressed (or even it vanishes by any chance) a relevant Page curve may not be recovered. However, we will show that this is not the case.

{
Two remarks are in order. As elucidated thus far, it is inadequate to account for finite tension on the brane in the presence of electric/magnetic charges. This implies a requisite for further investigation to elucidate the influence of tension beyond neutral black holes. Nevertheless, within the scope of this paper, we explore the scenario of zero tension ($\theta=\pi/2$) as a tensionless ``limit" ($\theta\rightarrow\pi/2$), i.e., a case of small tension. This can be justified by the fact that the physics, particularly with regard to the Page curve, remains unaltered in scenarios of small tension compared to those of zero tension, e.g., \cite{Ling:2020laa}.

Moreover, with the advent of a novel method for examining tension in the presence of finite charge, the exploration of the opposite scenario--the large tension case (or very small values of $\theta$)--will also become viable.\footnote{This perspective may be more relevant in the context of the $d$-dimensional effective theory of gravity and matter through the lens of Randall-Sundrum.}\\}

In summary, within doubly holographic theories for dyonic black holes, there can be two Neumann boundary conditions imposed on the Planck brane: \eqref{BDRYT2}.
The former one gives the relation between the tension and the angle of the Planck brane: \eqref{TAR}, i.e., given strength of the tension $\alpha$, such a relation determines the location (or the angle $\theta$) of the Planck brane or vice versa.
On the other hand, the latter produces the ratio between the density and magnetic field, which is inversely proportional to topological coefficients: \eqref{RATD2}.

\subsection{The quantum extremal surface}

The entanglement entropy of the Hawking radiation (or equivalently of the black holes; recall that the entire system is in the pure state) can be measured by the quantum extremal surfaces~\cite{Penington:2019npb,Almheiri:2019hni,Almheiri:2019psf,Almheiri:2019psy,Penington:2019kki}. 
One can have two kinds of the quantum extremal surfaces: (I) the connected surface; (II) the disconnected surface. For instance, see the left figure in Fig. \ref{SK223FIG}.

In the doubly holographic framework, as demonstrated in the introduction, the quantum extremal surfaces can be equivalently described by the RT/HRT surface in one-higher dimensions~\cite{Almheiri:2019hni,Almheiri:2019psy,Ling:2020laa}: see the right figure in Fig. \ref{SK223FIG} where the connected surface is promoted into the Hartman-Maldacena surface (HM surface)~\cite{Hartman:2013qma}, while the disconnected surface is into the island surface.
In other words, using the doubly holography, one can simply evaluate the entanglement entropy of the radiation by 
\begin{align}\label{ISFO2}
\begin{split}
S_{\text{R}} = \min_{\text{I}} \left[ \frac{\operatorname{Area}\left( \Gamma_{\text{I} \cup \text{R}} \right) }{4 G_{N}} \right] \,,
\end{split}
\end{align}
where $\Gamma_{\text{I} \cup \text{R}}$ is the standard co-dimension two HRT surface in the bulk, which is corresponding to the HM surface ($\Gamma_{\text{HM}}$) or island surface ($\Gamma_{\text{Is}}$). In the next section, we give the details of how to obtain both the HM surface and the island surface.

Note that the HM surface (blue solid line) is anchored on the left/right baths when the island is absent and passes through the black hole horizon, while the island surface (orange solid line) is anchored on the Planck brane containing the island, which is outside the horizon.
This implies that the entropy computed using the HM surface can exhibit the time evolution because of the stretching of space inside the horizon as described in \cite{Hartman:2013qma}, while the one from the island surface does not since it does not penetrate the horizon (i.e., the entropy from the island surface does not get affected by the stretching of space inside the horizon so it is time-independent).

Thus, one can expect that the initial rise of the entanglement entropy in the Page curve is described by the HM surface and the saturation of the entanglement entropy at late times is by the island surface. We will show that this is the case by the explicit calculations.

Alternatively, one can also say that the Page curve cannot be achieved if the island surface dominates over the HM surface at $t=0$, i.e., the Page curve is being saturated at $t=0$ (it is equivalent to say the Page time $t_P=0$).
It is also worth noticing that this issue may arise in the context of doubly holographic framework, which was noticed in \cite{Geng:2020qvw} first and elaborated further in \cite{Geng:2020fxl}. See also \cite{Geng:2021mic} for the related topic: the constant entropy belt.
However, it is shown \cite{Ling:2020laa} that the resolution of it is simply to take the large value of $x_b$ (i.e., moving the end point of the radiation region away from the Planck brane).\footnote{See also \cite{Ling:2020laa} for the related treatment at $t=0$ using DGP term when the tension is taken into account.}

Furthermore, in this paper, we show that this resolution (taking large $x_b$) can also be further implemented to find the double Bekenstein-Hawking entropy in the Page curve at late times.

%
\section{Some formalism for extremal surfaces}\label{sec3la}

In this section, we present the holographic calculation of two extremal surfaces (HM surface $\Gamma_{\text{HM}}$ and island surface $\Gamma_{\text{Is}}$) in detail. In particular, we present not only the holographic formula of both surfaces at $t=0$, but also the time-dependent HM surface.

For this purpose, we consider a general asymptotically $\text{AdS}_{d+1}$ metric as
\begin{equation}\label{GENMET}
\dd s^2 = \frac{L^{2}}{z^2} \left[-f_{0}(z) \, \dd t^2 \,+\,  f_{1}(z) \, \dd x^2_{i} \,+\, f_{2}(z) \, \dd z^2  \right] \,,
\end{equation}
where $i = 1,\dots, d-1$ and $f_{0}(z)$, $f_{1}(z)$, and $f_{2}(z)$ are approaching to $1$ at the AdS boundary ($z\rightarrow0$).

%
\subsection{The extremal surfaces at $t=0$}

Let us first discuss the extremal surfaces at fixed time $t=0$, i.e., the induced metric of it does not contain $f_0(z)$ of \eqref{GENMET}. {Recall that in this paper, we focus on the tensionless brane, for which its backreaction to the background geometry can be negligible.}\\

\noindent \textbf{Entanglement entropy from the island surface.}
Since the island surface (the solid orange line in Fig. \ref{SK223FIG}) is the standard Ryu-Takayanagi surface of the subsystem length $x_b$, the holographic entanglement entropy of the island surface can be simply obtained as 
\begin{align}\label{ISLAND1}
\begin{split}
S_{\text{Is}}(t=0) &:= \frac{\text{Area}(\Gamma_{\text{Is}})}{4 G_{N}} \\
&= \frac{2 L^{d-1} \Omega^{d-2}}{4 G_{N}} \int^{z_{b}}_{\epsilon} \frac{\dd z}{z^{d-1}} \,\, \frac{\sqrt{f_2(z) f_1^{d-2}(z)}}{\sqrt{1- \frac{f_1^{d-1}(z_{b}) z^{2d-2} }{f_1^{d-1}(z) z_{b}^{2d-2}}}} \,, 
\end{split}
\end{align}
where $\Omega_{d-2}$ is a volume of the $(d-2)$ spatial directions, $z_{b}$ is the deepest point of the minimal surface in the bulk (i.e., the orange point in Fig. \ref{SK223FIG}), and $\epsilon$ the UV cutoff.\footnote{{
The extremal surface for \eqref{ISLAND1} can be found by solving an extremization problem with a function $z(x_1)$ in which the geometric symmetry in the $x_1$ direction produces a closed form for $z'(x_1)$ in terms of a conserved quantity.}}
One can also find the relation between $x_{b}$ and $z_{b}$ from the minimization of the area as
\begin{align}\label{ISLAND3}
\begin{split}
x_{b} = \int_{0}^{z_{b}}  \dd \alpha \, \frac{\sqrt{\frac{f_{2}(\alpha)}{f_{1}(\alpha)}}}{\sqrt{\frac{f_{1}^{d-1}(\alpha) \, z_{b}^{2d-2}}{f_{1}^{d-1}(z_{b}) \, \alpha^{2d-2}}} -1 }  \,.
\end{split}
\end{align}

Note that when the metric has a simple form as
\begin{align}\label{YILINGRESULT}
\begin{split}
f_{0}(z) = f(z) \,,\quad f_{1}(z) = 1 \,,\quad f_{2}(z) =  \frac{1}{f(z)} \,,
\end{split}
\end{align}
where $f(z)$ is the emblackening factor, all formulae \eqref{ISLAND1}-\eqref{ISLAND3} reduce to results in \cite{Ling:2020laa} of which the Planck brane is tensionless. Note also that our formulae are consistent with the usual holographic entanglement entropy of a strip subsystem when the subsystem size $\ell = 2 x_{b}$, for instance, see \cite{Ben-Ami:2016qex}.\\

\noindent \textbf{Entanglement entropy from the Hartman-Maldacena surface.}
The holographic entanglement entropy of the Hartman-Maldacena surface (HM surface; the solid blue line in Fig. \ref{SK223FIG})~\cite{Hartman:2013qma} at $t=0$ can be obtained as
\begin{align}\label{TRIVIAL1}
\begin{split}
S_{\text{HM}}(t=0) &:= \frac{\text{Area}(\Gamma_{\text{HM}})}{4 G_{N}} \\
&= \frac{2 L^{d-1} \Omega^{d-2}}{4 G_{N}} \int^{z_{h}}_{\epsilon} \frac{\dd z}{z^{d-1}} \,\, \sqrt{f_{2}(z) f_{1}^{d-2}(z)} \,,
\end{split}
\end{align}
where $z_h$ is the horizon radius and it is consistent with \cite{Ling:2020laa} when the simple metric \eqref{YILINGRESULT} is chosen.\footnote{{
The extremal surface corresponding to \eqref{TRIVIAL1} can be determined with the symmetry in $(t, x_i)$, identifying it as the surface that fall straight into the bulk.}}
Notice that, when we consider the one-higher dimensional object of \eqref{TRIVIAL1}, we obtain the holographic complexity formula (via complexity=volume conjecture) \cite{Carmi:2017jqz,Yang:2019gce} as expected: recall that there is one dimensional difference between the area (i.e., entanglement entropy) and the volume (i.e., complexity).

In the next section, we will numerically compute the time-dependent $S_{\text{HM}}$ in the Eddington-Finkelstein coordinate and show that the numerical result at $t=0$ is consistent with the analytic expression \eqref{TRIVIAL1}.\\

\noindent \textbf{The UV-finite holographic entanglement entropy.}
Using all the holographic formulae of the entanglement entropy of the Hawking radiation above, \eqref{ISLAND1} and \eqref{TRIVIAL1}, one can study which entropy is dominated at $t=0$, for instance, if $S_{\text{Is}} > S_{\text{HM}}$, $S_{\text{R}}$ in \eqref{ISFO2} corresponds to $S_{\text{HM}}$.

However, since both \eqref{ISLAND1} and \eqref{TRIVIAL1} are UV-divergent quantities, we first need to regularize the entanglement entropy. Note that both entropies have the same structure of UV-divergence as 
\begin{align} \label{DIVTERM}
S_{\text{Is or HM}} \,\approx\, \frac{\#}{\epsilon^{d-2}} \,+\, S_{\text{Is or HM}}^{\text{Finite}} \,,
\end{align}
where the divergent term, $1/\epsilon^{d-2}$, originates from the contribution near the AdS boundary.

There may be several ways to regulate the entropy given in the literature. For instance, one can simply omit the divergence term by hand and study the remained finite piece. Another way is to study the difference between $S_{\text{Is or HM}}$ and the one from a pure AdS geometry ($S_{\text{AdS}}$) usually interpreted as the entanglement entropy of the ground state of the CFT: note that in this way the finite piece of $S_{\text{Is or HM}}$ can be slightly varied by the finite piece of $S_{\text{AdS}}$.

Nevertheless, the other type of regularization has been implemented for the study of the Page curve in doubly holography~~\cite{Almheiri:2019psy, Ling:2020laa} as
\begin{align} \label{FORMU}
\begin{split}
\Delta S_{\text{HM}}(t)  \,&:=\,S_{\text{HM}}(t) \,-\, S_{\text{HM}}(t=0) \\
&\,=\, S^{\text{Finite}}_{\text{HM}}(t) \,-\, S^{\text{Finite}}_{\text{HM}}(t=0) \,, \\
\\
\Delta S_{\text{Is}}(t) \,&:=\,S_{\text{Is}}(t) \,-\, S_{\text{HM}}(t=0) \\
&\,=\, S^{\text{Finite}}_{\text{Is}}(t) \,-\, S^{\text{Finite}}_{\text{HM}}(t=0) \,, 
\end{split}
\end{align}
where \eqref{TRIVIAL1} is used for the regularization. Note that this kind of regularization can be justified for the purpose of the Page curve: the time evolution of the entanglement entropy in which its growth is of main interest.

Furthermore, \eqref{FORMU} can also be useful to discuss if the Page curve can be achieved or not at $t=0$. For instance, based on the explanation described below \eqref{ISFO2}, one needs to find $\Delta S_{\text{Is}}(t=0)>0$ in order to obtain the Page curve at a finite Page time: otherwise, the entanglement entropy is already saturated by the island surface at $t=0$.
Strictly speaking, $\Delta S_{\text{Is}}(t)$ is a time-independent quantity since the island surface cannot penetrate the horizon unlike the HM surface~\cite{Hartman:2013qma}, i.e., $\Delta S_{\text{Is}}(t) = \Delta S_{\text{Is}}(t=0)$ for all time. We will discuss more on this point when we display the Page curve of dyonic black holes.\\

\begin{figure}[]
\centering
     {\includegraphics[width=8.0cm]{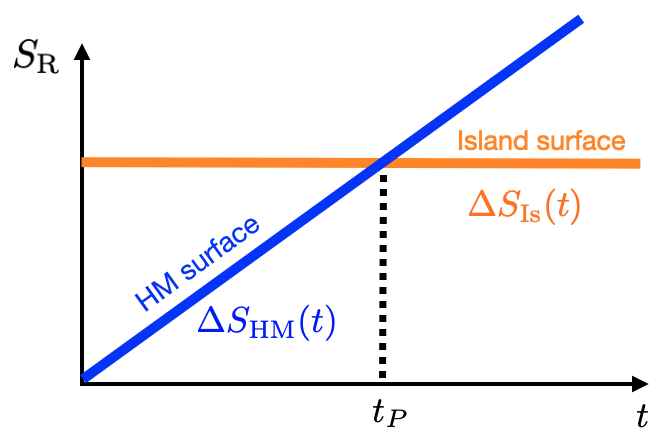} \label{}}
 \caption{A sketch of the Page curve of the entanglement entropy of the Hawking radiation from \eqref{OVERALLPIC} where the HM surface is dominant before the Page time $t_P$ and the entropy is saturated after $t_P$ by the island surface.}\label{PCC}
\end{figure}

In summary, following the doubly holographic theories~\cite{Almheiri:2019psy, Ling:2020laa} for higher dimensional black holes, we consider the UV-finite entanglement entropy of the Hawking radiation \eqref{FORMU} in order to describe the Page curve as
\begin{align}\label{OVERALLPIC}
S_{\text{R}} \,=\,
\begin{cases}
\Delta S_{\text{HM}}(t) \,,    \quad       (t < t_{P})     \\
\Delta S_{\text{Is}}(t)  \,,      \quad\,\,\,\,     (t \geqslant t_{P})    
\end{cases}
\end{align}
where $t_P$ denotes the Page time. For instance, see Fig. \ref{PCC}.
\begin{figure*}[]
 \centering
     {\includegraphics[width=15cm]{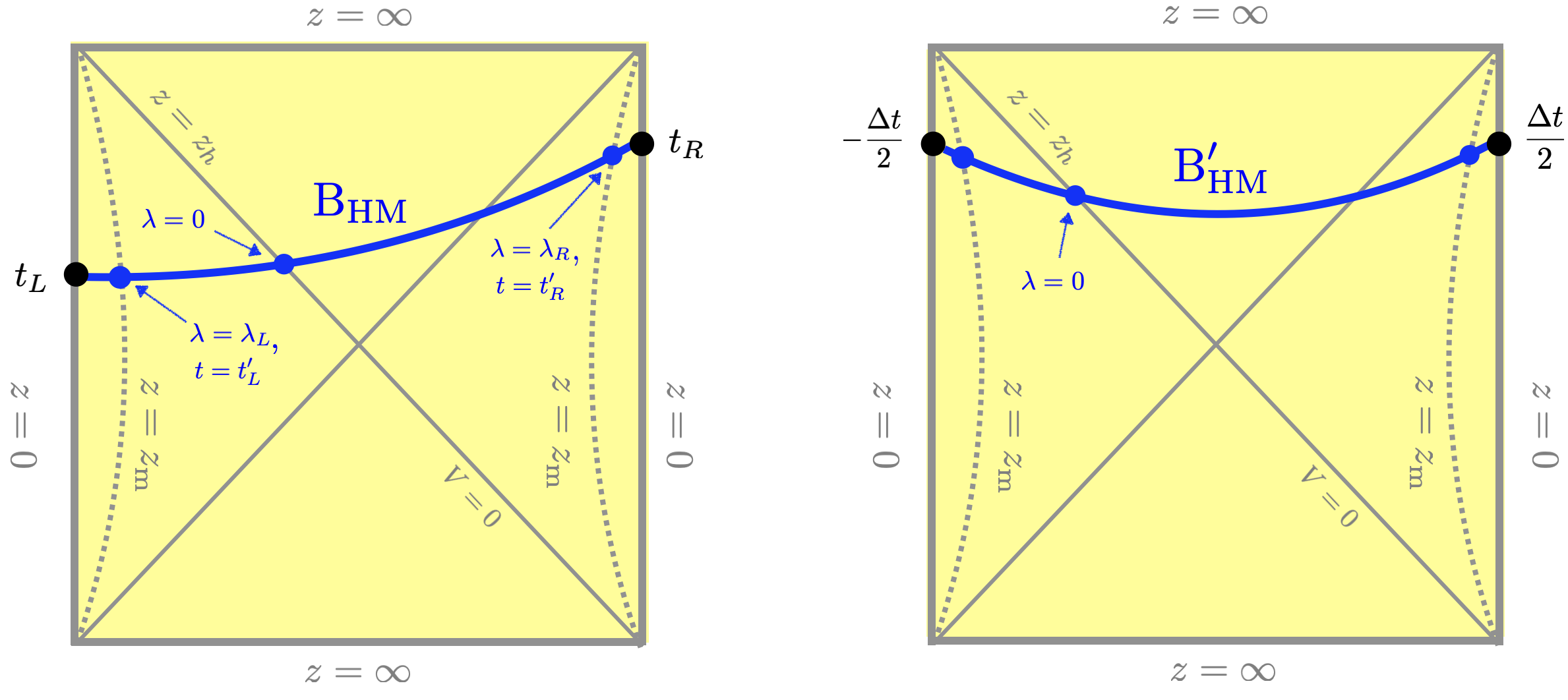} \label{}}
\caption{A sketch of the time-dependent HM surface. Here $t_L$ and $t_R$ denote for times at each boundary and $z_h$ the horizon radius. $\lambda$ is introduced to parametrize $z(\lambda)$ and $V(\lambda)$.
$\text{B}_{\text{HM}}$ is the schematic representation of a {bulk surface} connecting $t_L$ and $t_R$ in the left figure, while $\text{B}_{\text{HM}}'$ is the one connecting $-\frac{\Delta t}{2}$ and $\frac{\Delta t}{2}$ in the right figure.} \label{penrosefig}
\end{figure*}
%

%
\subsection{Time-dependent Hartman-Maldacena surface}\label{sec32}

Next, we provide the detailed methodology to compute the time-dependent HM surface, $\Gamma_{\text{HM}} (t)$, which leads to obtaining $\Delta S_{\text{HM}}(t)$.
When we study the time evolution of the HM surface, we are essentially moving a {bulk surface} ($
\text{B}_{\text{HM}}$ or $\text{B}_{\text{HM}}'$) forwards in time ($t_{L}, t_{R}$) on both sides: see Fig. \ref{penrosefig}.
Note that the HM surface is moving in a time direction perpendicular to the right figure of Fig. \ref{SK223FIG}.

In what follows, we consider the metric ansatz \eqref{GENMET} in order to generalize the formalism beyond the simple setup \eqref{YILINGRESULT}, given in previous literature.
It is also worth noticing that the metric of interest has time translational symmetry. In other words, the final result depends only on the combination 
\begin{equation} \label{Dtrl}
\Delta t := t_{R} - t_{L} \,,
\end{equation}
rather than each of the boundary times ($t_L$ or $t_R$). This is due to the invariance of the system under the shift as $t_L \rightarrow t_L - \delta t$ and $t_R \rightarrow t_R - \delta t$, {see~\cite{Carmi:2017jqz}}.
Therefore, one can choose the symmetric configuration with time 
\begin{equation} \label{}
t_L := -\frac{\Delta t}{2} \,, \qquad t_R := \frac{\Delta t}{2} \,,
\end{equation}
in order to study the time evolution of the HM surface: see the right figure in Fig. \ref{penrosefig}.\\

\noindent \textbf{Reparametrization of the HM surface.}
In order to discuss the time-dependent HM surface $\Gamma_{\text{HM}}(t)$, it is convenient to rewrite the metric \eqref{GENMET} using the null coordinate $V(t, z)$ 
\begin{align}\label{Nullcoordi}
\begin{split}
V(t,z)  &=  e^{\bar{\beta} \, v(t,z)}  =  e^{\bar{\beta} \, (t - z^*(z))} \,, \\
z^*(z)  &=  \int_{0}^{z} \sqrt{\frac{f_2(\tilde{z})}{f_0(\tilde{z})}} \td \tilde{z} \,,
\end{split}
\end{align}
where $v(t,z)$ is the infalling Eddington-Finkelstein coordinate, and $\bar{\beta}$ will be determined by solving the equation of motion near the horizon: see \eqref{betaresult}.

In this null coordinate, the metric \eqref{GENMET} reads 
\begin{align}\label{Nullmetric}
\begin{split}
\td s^{2} &=  \frac{L^2}{z^2} \biggl\{ - f_0(z)  \td v^{2}  - 2\sqrt{f_0(z) f_2(z)}  \td v \td z  +  f_1(z) \sum_{i=1}^{d-1}\td x^2_i \biggr\} \,,\\
&=  \frac{L^2}{z^2} \biggl\{ - \frac{f_0(z)}{\bar{\beta}^2V^2} \td V^{2} - \frac{2 \sqrt{f_0(z) f_2(z)}}{\bar{\beta} V}\td V \td z \\
&\qquad\quad + f_1(z) \sum_{i=1}^{d-1}\td x^2_i \biggr\} \,.
\end{split}
\end{align}
Furthermore, one can find the induced metric of the {bulk surfaces} (solid blue curves) in Fig. \ref{penrosefig}, which can be parametrized by $\lambda$ as
\begin{widetext}
\begin{align}\label{Nullmetric}
\begin{split}
	\td s_{\text{HM}}^{2} :=  \frac{L^2}{z(\lambda)^2} \Bigg[ \left(- \frac{f_0(z(\lambda))}{\bar{\beta}^2V(\lambda)^2}V'(\lambda)^{2}  -  \frac{2\sqrt{f_0(z(\lambda)) f_2(z(\lambda))}}{\bar{\beta} V(\lambda)}V'(\lambda)z'(\lambda) \right) \td \lambda^2  +  f_1(z(\lambda)) \sum_{i=2}^{d-1}\td x^2_i \Bigg] \,.
\end{split}
\end{align}
\end{widetext}

Then, using this induced metric, we are led to find the area of the bulk surface, $\text{Area}(\Gamma_{\text{HM}})$, as
\begin{align}\label{Lag1Vol1}
\begin{split}
\text{Area}(\Gamma_{\text{HM}}) = L^{d-1} \, \Omega^{d-2}  \int \mathcal{L}_1\, \td \lambda \,,
\end{split}
\end{align}
where 
\begin{align}\label{}
\begin{split}
\mathcal{L}_1 := \, & \frac{f_1(z(\lambda))^{\frac{d-2}{2}}}{z(\lambda)^{d-1}} \biggl\{ - \frac{f_0(z(\lambda))}{\bar{\beta}^2V(\lambda)^2} V'(\lambda)^{2}  \\ 
& - \frac{2 \sqrt{f_0(z(\lambda)) f_2(z(\lambda))} }{\bar{\beta} V(\lambda)} V'(\lambda)z'(\lambda)  \biggr\}^{1/2}   \,,
\end{split}
\end{align}
and $\Omega^{d-2}$ is the volume of the spatial geometry as in \eqref{ISLAND1}.
In addition, one can also find that we only have a single Euler-Lagrangian equation from \eqref{Lag1Vol1}
\begin{widetext}
\begin{align}\label{Lag1eom}
\begin{split}
& -  \frac{z''}{z'}  + \frac{V''}{V'}  + \frac{1}{2}\left[ \frac{4(d-1)}{z} - \frac{f_0'(z)}{f_0(z)} - \frac{2(d-2)f_1'(z)}{f_1(z)} - \frac{f_2'(z)}{f_2(z)} \right] z' \\
& \quad + \frac{2(d-1)f_0(z) f_1(z) -z f_0'(z) f_1(z) - (d-2) z f_0(z) f_1'(z) }{2\bar{\beta}^2 z z' V^2 f_1(z) f_2(z)} V'^{2} \\
& \quad + \frac{6(d-1)f_0(z) f_1(z) - 2\bar{\beta} z f_1(z)\sqrt{f_0(z) f_2(z)} -3 z f_0'(z) f_1(z) - 3(d-2) z f_0(z) f_1'(z) }{2\bar{\beta} V z f_1(z) \sqrt{f_0(z) f_2(z)}} V' = 0 \,.
\end{split}
\end{align}
\end{widetext}
However, it would be problematic since we have two independent fields ($z(\lambda), V(\lambda)$). In order to resolve this, one can introduce the auxiliary field $\varepsilon(\lambda)$ to \eqref{Lag1Vol1}
\begin{align}\label{Lag2eom}
\begin{split}
\text{Area}(\Gamma_{\text{HM}})  &\,=\,  L^{d-1} \, \Omega^{d-2} \int \left( \frac{1}{\varepsilon(\lambda)}\mathcal{L}_1^2 + \varepsilon(\lambda) \right) \td \lambda \\
  &\,=:\,  L^{d-1} \, \Omega^{d-2} \int \mathcal{L}_2\, \td \lambda\,,
\end{split}
\end{align}
from which, we obtain three Euler-Lagrangian equations as
\begin{widetext}
\begin{align}\label{EL2}
\begin{split}
	& \frac{\bar{\beta} f_2(z)}{\sqrt{f_0(z) f_2(z)}}z'' - \frac{\bar{\beta} f_2(z) \varepsilon'}{\varepsilon \sqrt{f_0(z) f_2(z)}} z' + \frac{V''}{V} - \frac{V'^{2}}{V^2} -\left[ \frac{\varepsilon'}{\varepsilon} - \left( \frac{2-2d}{z} + \frac{f_0'(z)}{f_0(z)} + \frac{(d-2)f_1'(z)}{f_1(z)} \right)z' \right]\frac{V'}{V}  \\
	&\quad  -\Bigg[ f_0(z) \Big\{ f_1(z) \left(4(d-1)f_2(z) - z f_2'(z)\right) - 2(d-2) z  f_1'(z) f_2(z) \Big\}  - z f_0'(z) f_1(z) f_2(z)  \Bigg] \frac{\bar{\beta} z'^2 f_2(z)}{2 z f_1(z) \left( f_0(z) f_2(z) \right)^{3/2}}  = 0 \,,  \\ \\
	& V'' - \Bigg[ f_1(z) \sqrt{f_0(z) f_2(z)} z f_0'(z)  + f_0(z) \Big\{ 2f_1(z) \left( \bar{\beta} f_2(z) z -(d-1)\sqrt{f_0(z) f_2(z)} \right) \\ 
	& \qquad\quad + (d-2)\sqrt{f_0(z) f_2(z)} z f_1'(z) \Big\}  \Bigg] \frac{V'^{2}}{2\bar{\beta} f_0(z) f_1(z) f_2(z) V z} - \frac{\varepsilon' V'}{\varepsilon} = 0 \,,  \\ \\
	& 1 + \frac{z^{-2(d-1)}}{\varepsilon^2}\left( \frac{2 f_1(z)^{d-2} \sqrt{f_0(z) f_2(z)}  V' z'}{\bar{\beta} V} + \frac{ f_0(z) f_1(z)^{d-2} V'^2}{\bar{\beta}^2 V^2} \right) = 1- \frac{\mathcal{L}_1^2}{\varepsilon^2} =0 \,.
\end{split}
\end{align}
\end{widetext}
Furthermore, one can also check that only two of these three equations are independent: here, we choose the second and third equations of motion.\\

\noindent \textbf{Equations of motion for the area of HM surface.}
Hereafter we take the auxiliary field $\varepsilon(\lambda)$=1 in order to recover the original variational problem.
Then, two independent equations of motion read
\begin{widetext}
\begin{align}\label{EOMGEN}
\begin{split}
	& V'' - \Bigg[ f_1(z) \sqrt{f_0(z) f_2(z)} z f_0'(z)  + f_0(z) \Big\{ 2f_1(z) \left( \bar{\beta} f_2(z) z -(d-1)\sqrt{f_0(z) f_2(z)} \right) \\ 
	& \qquad\quad + (d-2)\sqrt{f_0(z) f_2(z)} z f_1'(z) \Big\}  \Bigg] \frac{V'^{2}}{2\bar{\beta} f_0(z) f_1(z) f_2(z) V z}  = 0 \,, \\ \\
	& 1 + z^{-2(d-1)} \left( \frac{2 f_1(z)^{d-2} \sqrt{f_0(z) f_2(z)}  V' z'}{\bar{\beta} V} + \frac{ f_0(z) f_1(z)^{d-2} V'^2}{\bar{\beta}^2 V^2} \right) = 1- \mathcal{L}_1^2 =0 \,.
\end{split}
\end{align}
\end{widetext}
Then, we solve these equations of motion numerically by the shooting method where we perform the shooting from the horizon to the two boundaries. 

Note that the horizon is located at $\lambda=0$, i.e., $(V(0)\,, z(0)) = (0\,, z_h)$: see the blue dot for $\lambda=0$ in Fig. \ref{penrosefig}. 
In addition, one can find the series solutions near the horizon ($\lambda = 0$) as 
%
\begin{align}\label{bcbtz}
\begin{split}
	 V &= V_{(1)} \, \lambda + \biggl( \frac{d-1}{z_h} - \frac{(d-2) f_1'(z_h)}{2 f_1(z_h)} \\
	 & \qquad\qquad\quad + \frac{\chi'(z_h)}{4\chi(z_h)} - \frac{f_0''(z_h)}{2f_0(z_h)} \biggr) z_{(1)} V_{(1)} \, \lambda^2 + \cdots \,,\\
\\
	z &= z_h + z_{(1)} \, \lambda + \biggl\{ \frac{z_h^{2d-2} f_0'(z_h) }{4 \chi(z_h) f_1(z_h)^{d-2}} \\
	& \qquad + \left( \frac{d-1}{z_h} - \frac{(d-2) f_1'(z_h)}{2 f_1(z_h)} - \frac{\chi'(z_h)}{4 \chi(z_h)} \right) z_{(1)}^{2} \biggr\} \, \lambda^2 + \cdots \,,
\end{split}
\end{align}
%
where we have two-independent shooting parameters ($V_{(1)}$, $z_{(1)}$).
Solving the equations of motion at the leading order, one can also find the value of $\bar{\beta}$ introduced in \eqref{Nullcoordi} as
\begin{equation}\label{betaresult}
\bar{\beta} =  -\frac{f_0'(z_{h})}{2\sqrt{\chi (z_h)}} \,, \qquad \chi(z_h) := f_0 (z_h) \, f_2 (z_h) \,.
\end{equation}
One can easily check that $\bar{\beta} = 2 \pi T$ as expected from the structure of the null coordinate.

In order for the numerical calculations, we further introduce $\tilde{z} = z/z_{h}$ without loss of generality, i.e., it is convenient to set $z_h=1$ for numerics.
Then, given value of the shooting parameters ($V_{(1)}$, $z_{(1)}$), one can solve \eqref{EOMGEN} and find the corresponding numerical solutions ($\tilde{z}(\lambda), V(\lambda)$).

Furthermore, one can determine the value of ($\lambda_{R}$\,, $\lambda_{L}$) from one of the numerical solution, $\tilde{z}(\lambda)$, at the given cut-off $\tilde{z}_{\text{m}}$ (see Fig. \ref{penrosefig}) as
\begin{align}\label{lambdaval}
\begin{split}
   \tilde{z}(\lambda_{R}) = \tilde{z}_{\text{m}}   \,, \qquad    \tilde{z}(\lambda_{L}) = \tilde{z}_{\text{m}}   \,,
\end{split}
\end{align}
where $\lambda_R>0$ and $\lambda_L<0$.

Therefore, once ($\lambda_{R}$\,, $\lambda_{L}$) are evaluated from the numerics, we are finally led to the computation of the area \eqref{Lag1Vol1} as 
\begin{align}\label{time111}
\begin{split}
\text{Area}(\Gamma_{\text{HM}})  &= L^{d-1}\, \Omega^{d-2} \int \mathcal{L}_1\, \td \lambda  \\
&=  L^{d-1} \, \Omega^{d-2} \int \, \td \lambda \\
&=  L^{d-1} \, \Omega^{d-2}(\lambda_{R} - \lambda_{L}) \,,
\end{split}
\end{align}
since we can use the fact $\mathcal{L}_1=1$ from the second equation of motion in \eqref{EOMGEN}.\footnote{
{$\mathcal{L}_1=1$ in \eqref{Lag1Vol1} can also be associated with the reparametrization invariance giving a single equation of motion \eqref{Lag1eom}.}} In other words, the area can be simply obtained by the difference $\lambda_{R} - \lambda_{L}$.

In addition, using all the numerical solutions ($\tilde{z}(\lambda), V(\lambda)$) together with the definition of the null coordinate \eqref{Nullcoordi}, we can also find the times at the boundary cut-off $\tilde{z}_{\text{m}}$ as
\begin{align}\label{tLRval}
\begin{split}
t_{R}^{'}  &:= t(\lambda_{R}) =  \frac{\log({|V(\lambda_{R})|})}{\bar{\beta}} + z^{*}(\lambda_{R})  \,, \\
t_{L}^{'}   &:= t(\lambda_{L})  =  \frac{\log({|V(\lambda_{L})}|)}{\bar{\beta}} + z^{*}(\lambda_{L}) \,,
\end{split}
\end{align}
which leads to find \eqref{Dtrl}
\begin{equation}
\Delta t =  t_{R}^{'} - t_{L}^{'}  =  \frac{1}{\bar{\beta}}\Big[\log({|V(\lambda_{R})|}) - \log({|V(\lambda_{L})|})\Big] \,,
\end{equation}
where we take $z^{*}(\lambda_{R}) - z^{*}(\lambda_{L}) {= 0}$ by \eqref{Nullcoordi} and \eqref{lambdaval}.\footnote{{Note that ($t_{R}^{'}, t_{L}^{'}$) are evaluated at finite $\tilde{z}_{\text{m}}$, which can be identical with the boundary times ($t_{R}, t_{L}$) at the AdS boundary.}} Strictly speaking, $\Delta t$ is the value in the $\tilde{z}_\text{m} \rightarrow 0$ limit, however $\Delta t$ can be defined at $\tilde{z}_\text{m} = 10^{-2}$ for the numerical calculation where $z^{*}(\lambda_{R}) - z^{*}(\lambda_{L}) = 0$ is valid.\\

In summary, by solving the equations of motion \eqref{EOMGEN}, we find the numerical solutions. Then, using the corresponding numerical solutions together \eqref{lambdaval}-\eqref{tLRval}, we evaluate the time-dependent HM surface, i.e., the time-dependent entanglement entropy as
\begin{align}\label{numemeth}
\begin{split}
     S_{\text{HM}}(t)  &= \frac{\text{Area}(\Gamma_{\text{HM}})}{4 G_{N}} = \frac{L^{d-1} \, \Omega^{d-2}}{4 G_{N}} \Big[ \lambda_{R}(\tilde{z}_{\text{m}}) - \lambda_{L}(\tilde{z}_{\text{m}}) \Big] \,, \\
     \Delta t &=  \frac{1}{\bar{\beta}}\left[\log\frac{V(\lambda_{R}(\tilde{z}_\text{m}))}{V(\lambda_{L}(\tilde{z}_\text{m}))}\right] \,.
\end{split}
\end{align}
For the sake of simplicity, we shall henceforth set the parameter $\frac{L^{d-1} \, \Omega^{d-2}}{4 G_{N}}$ to unity.

%
\section{Results of dyonic black holes}\label{sec4la}

In this section, implementing all the methods presented in the previous section, we study the Page curve of the dyonic black holes \eqref{TOTAC} within a doubly holographic setup.

\subsection{Holographic setup}
 
Using the simple metric ansatz \eqref{YILINGRESULT} together with the one for the gauge field as 
\begin{equation}\label{BGMET}
\begin{split}
A = A_t(z) \, \dd t -\frac{B}{2} y \,\dd x + \frac{B}{2} x \, \dd y \,,
\end{split}
\end{equation}
one can find the analytic background solution
\begin{equation}\label{BCF}
\begin{split}
 f(z) &\,= 1 - m_{0} z^3 \,+ \, \frac{\mu^2 z_{h}^2 + B^2 z_h^4}{4} \, \frac{z^4}{z_h^4} \,, \\
 m_{0} &\,= z_{h}^{-3} \left( 1 +  \frac{\mu^2 z_{h}^2 + B^2 z_{h}^4}{4} \right) \,, \quad
 A_{t}(z) \,= \mu \left( 1- \frac{z}{z_{h}} \right) \,,
\end{split}
\end{equation}
where $\mu$ is the chemical potential, $B$ magnetic field, and $z_{h}$ horizon radius.
Furthermore the various thermodynamic parameters including the temperature $T$, density $\rho$, and Bekenstein-Hawking entropy density $S_{\text{BH}}$ read
\begin{align}\label{HAWKINGT}
\begin{split}
 T = \frac{1}{4\pi} \left( \frac{3}{z_{h}} - \frac{B^2 z_{h}^3 + \mu^2 z_{h}}{4}  \right) \,, \,\,\, 
\rho = \frac{\mu}{z_h}  \,,  \,\,\,\, 
S_{\text{BH}} = \frac{4\pi}{z_{h}^2}  \,.
\end{split}
\end{align}
Here, we set the gravitational constant $16\pi G_N=1$ and AdS radius $L=1$ to avoid clutter.\footnote{{Together with $\frac{L^{d-1} \, \Omega^{d-2}}{4 G_{N}}=1$, this implies that we set $4\pi \Omega^{d-2}=1$.}}

In order for the numerical calculations, it is also convenient to introduce the tilde variables as 
\begin{align}\label{TV1}
\begin{split} 
\tilde{B} &:=  B \, z_h^2 \,,\quad \tilde{T} := T \, z_h = \frac{1}{4\pi} \left( 3 - \frac{\tilde{B}^2 + \tilde{\mu}^2}{4}  \right) \,, \\ 
  \tilde{\mu} &:= \mu \, z_h \,,\quad \,\,\,
\tilde{\rho} := \rho \, z_h^2 = \tilde{\mu}   \,,
\end{split}
\end{align}
which would be equivalent to set $z_h=1$.

In this paper, as the extension of the previous study~\cite{Ling:2020laa}, we also study the entanglement entropy \eqref{OVERALLPIC} at the fixed chemical potential. In other words, we aim to evaluate ${S_{\text{R}}}/{\mu}$ in terms of $\left( {T}/{\mu}\,, {B}/{\mu^2} \right)$.
In order for this, 
one needs to solve the following
\begin{align}\label{TV2}
\begin{split} 
\frac{T}{\mu} = \frac{\tilde{T}}{\tilde{\mu}} = \frac{1}{4\pi \tilde{\mu}} \left( 3 - \frac{\tilde{B}^2 + \tilde{\mu}^2}{4} \right) \,, \qquad \frac{B}{\mu^2} = \frac{\tilde{B}}{\tilde{\mu}^2}  \,.
\end{split}
\end{align}
and find the relation $\tilde{B} = \tilde{B} (T/\mu, B/\mu^2)$ and $\tilde{\mu} = \tilde{\mu} (T/\mu, B/\mu^2)$. Then, finally ${S_{\text{R}}}/{\mu}$ can be rewritten as ${S_{\text{R}}}/{\mu}\left( {T}/{\mu}\,, {B}/{\mu^2} \right)$.\footnote{{The intermediate step of rescaling with the horizon \eqref{TV1} can be useful for the numerical computations ($z_h=1$). One can directly find the rescaling with $\mu$ from the fact that black hole is invariant under $(z_h,\mu,B,T,\rho)\rightarrow(z_h\alpha, \mu/\alpha, B/\alpha^{2}, T/\alpha, \rho/\alpha^{2})$ together with $(t,z,x,y,\Omega) \rightarrow \alpha(t,z,x,y,\Omega/\alpha^{2})$ when $\alpha$ is a positive constant.}}\\

\noindent \textbf{Thermodynamic property in doubly holographic setup.}
As demonstrated in the section \ref{sec2la}, in the doubly holographic setup, the density and the magnetic field are no longer the independent parameters: \eqref{RATD2}. They are associated through the coefficients introduced on the Planck brane, $c_{pb}$ in \eqref{SNEQ}.
As we will show below, this further implies that at fixed chemical potential, $T/\mu$ is related to $B/\mu^2$ in the presence of $c_{pb}$.

\begin{figure*}[]
\centering
     {\includegraphics[width=7.0cm]{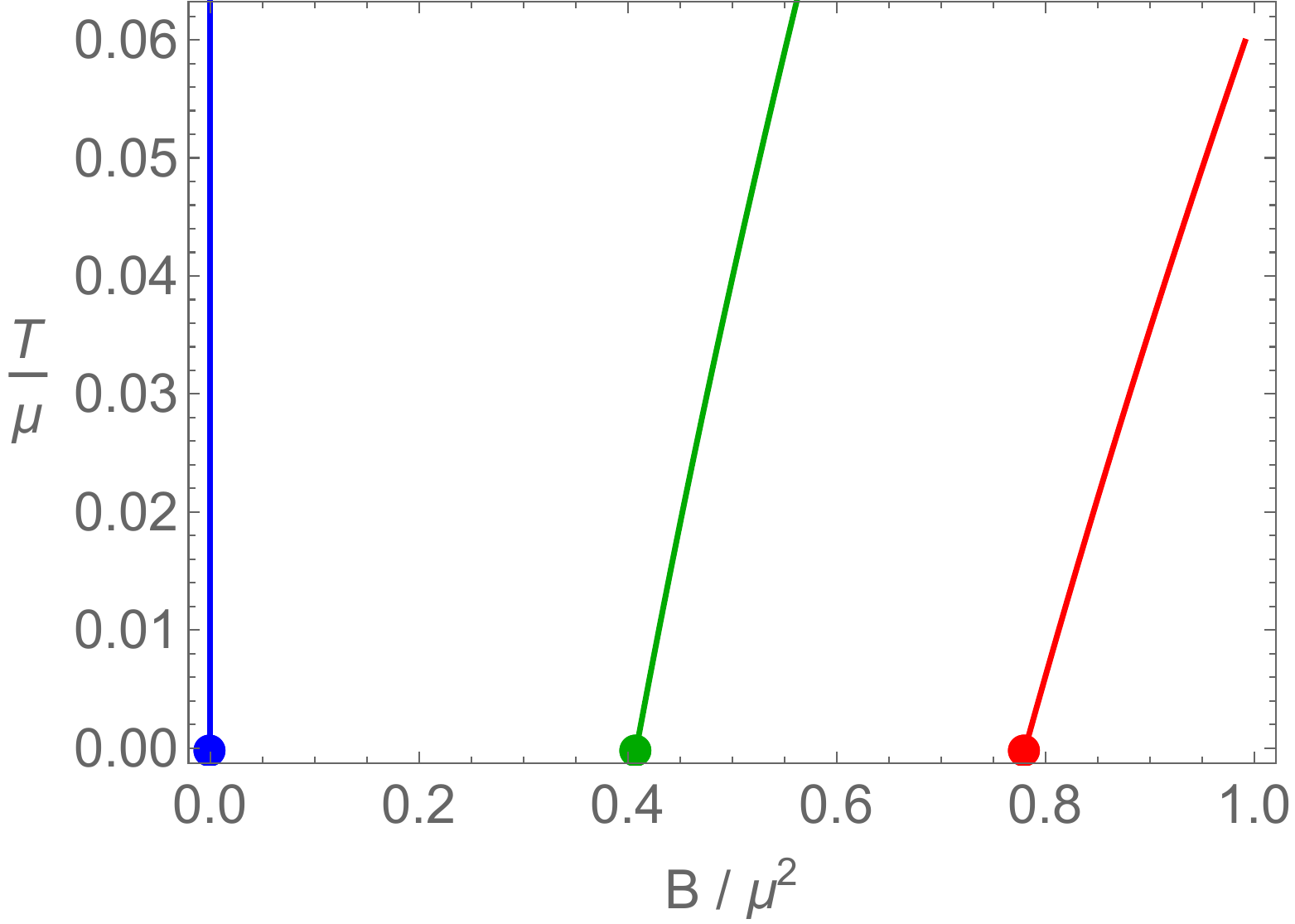} \label{}}
\qquad
     {\includegraphics[width=6.8cm]{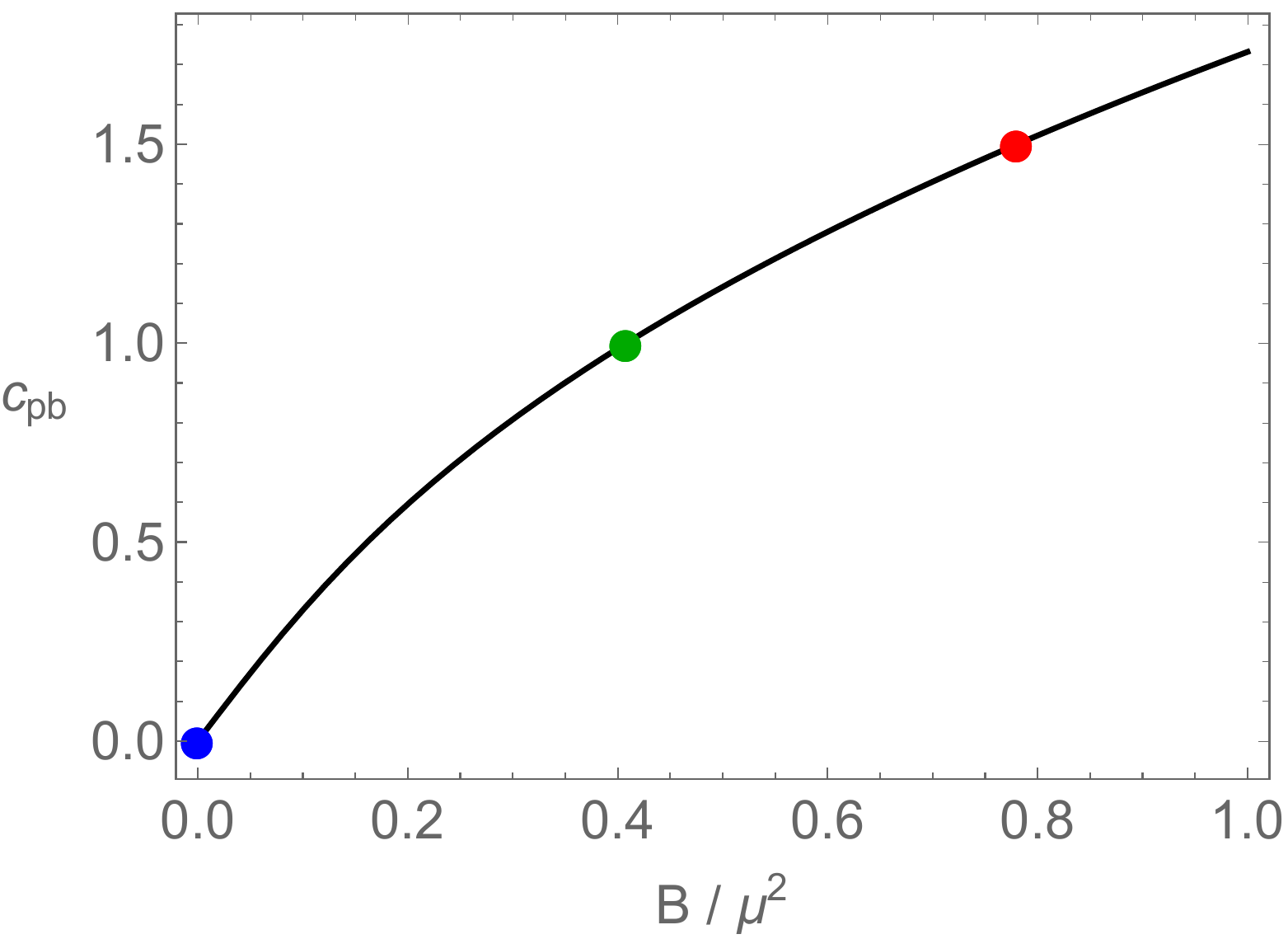} \label{}}
 \caption{Data with different value of $c_{pb} = 0, 1, 1.5$ (blue, green, red). The left figure shows $T$ vs. $B$ at given $c_{pb}$, while the right figure shows $c_{pb}$ vs. $B$ at given $T$. In particular, we set $T=0$ in the right figure where the black solid line is an analytic expression in \eqref{CHFOR}. Dots denote the same data in both figures.}\label{TBcfig}
\end{figure*}

Note that the relationship \eqref{RATD2} can be further expressed as a function of ($T/\mu\,, B/\mu^2$) using \eqref{TV1}-\eqref{TV2}:
\begin{align}\label{TDRCPB}
\begin{split}
\frac{1}{c_{pb}} \,=\, \frac{\rho}{B} \,=\, \frac{\tilde{\rho}}{\tilde{B}} \,=\, \frac{\tilde{\mu}}{\tilde{B}} \,=\, \mathcal{F} \left( \frac{T}{\mu}\,, \frac{B}{\mu^2} \right) \,.
\end{split}
\end{align}
One can find the explicit form of $\mathcal{F} \left( {T}/{\mu}\,, {B}/{\mu^2} \right)$. However, it is so complicated and not illuminating as well. Thus, we make a plot of it in Fig. \ref{TBcfig}.

In the left figure of Fig. \ref{TBcfig}, we display $T/\mu$ vs. $B/\mu^2$ at given values of $c_{pb}$. When $c_{pb}=0$ (blue data) corresponds to the setup in \cite{Ling:2020laa} where $B/\mu^2=0$ for all $T/\mu$. 
On the other hand, when $c_{pb}$ is non-vanishing one may have all $T/\mu$ after the certain (minimum) value of $B/\mu^2$. For instance, when $c_{pb}=1$ (green data) such a minimum magnetic field would be $B/\mu^2\approx 0.4$.

In the right figure of Fig. \ref{TBcfig}, we also show $c_{pb}$ vs. $B/\mu^2$ at given temperature $T/\mu=0$, where its analytical expression reads
\begin{align}\label{CHFOR}
\begin{split}
c_{pb}^2 \,(T=0) = \frac{\sqrt{1+48 (B/\mu^2)^2}-1}{2} \,, 
\end{split}
\end{align}
Note the right figure corresponds to the collection of all minimum values of  $B/\mu^2$ in the left figure, i.e., the dots indicate the same data in both figures.
Also notice that Fig. \ref{TBcfig} implies that at given $T$, one should consider a finite topological coefficient on the Planck brane in order to have a finite magnetic field.

\subsection{Entanglement entropy from HM surface}
Based on the thermodynamic relation \eqref{TDRCPB} including the topological coefficient, we study the aspects of two extremal surfaces of dyonic black holes: HM surface and island surface. 

\begin{figure*}[]
\centering
     {\includegraphics[width=6.8cm]{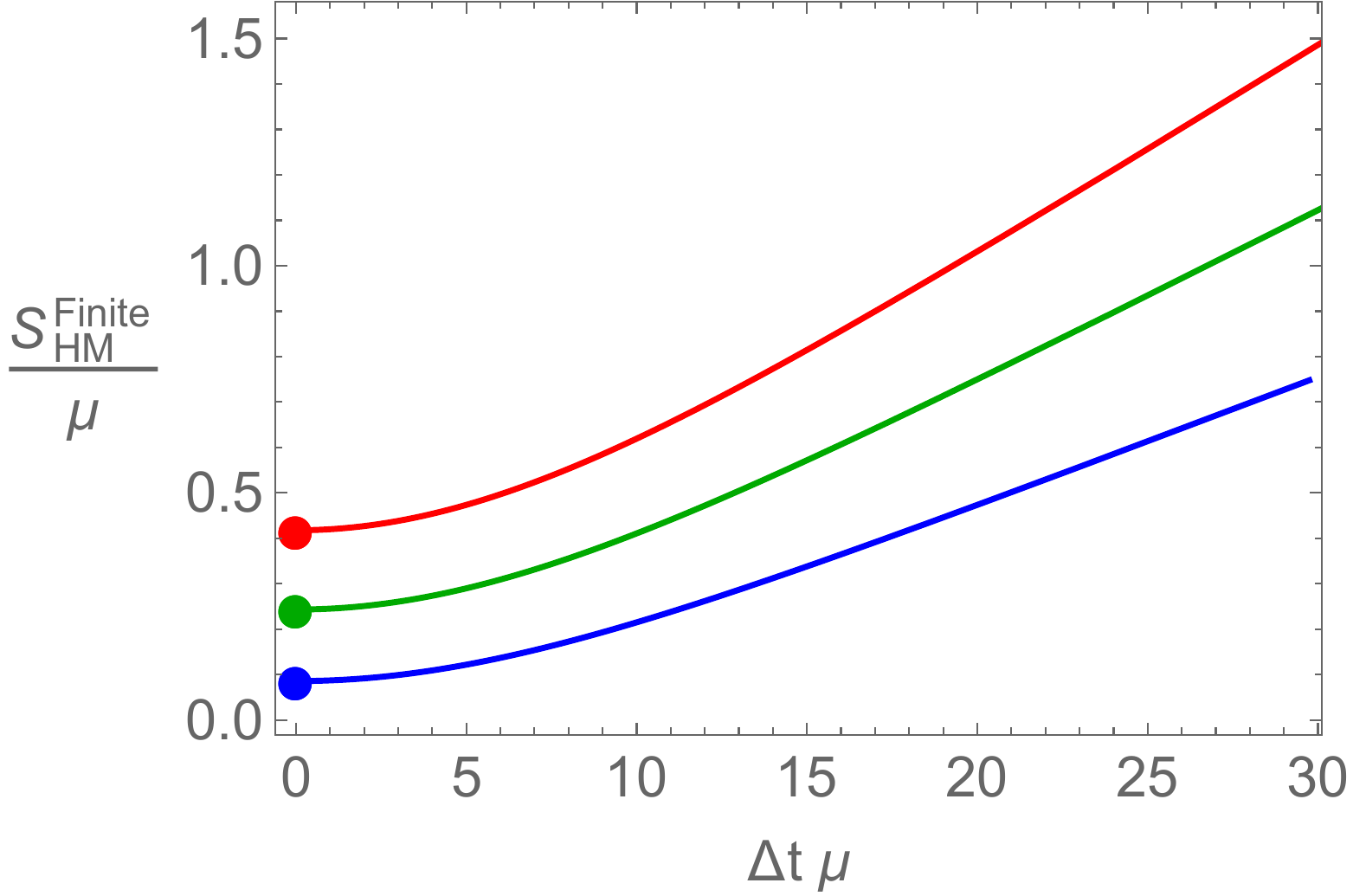} \label{}}
\qquad
     {\includegraphics[width=7.7cm]{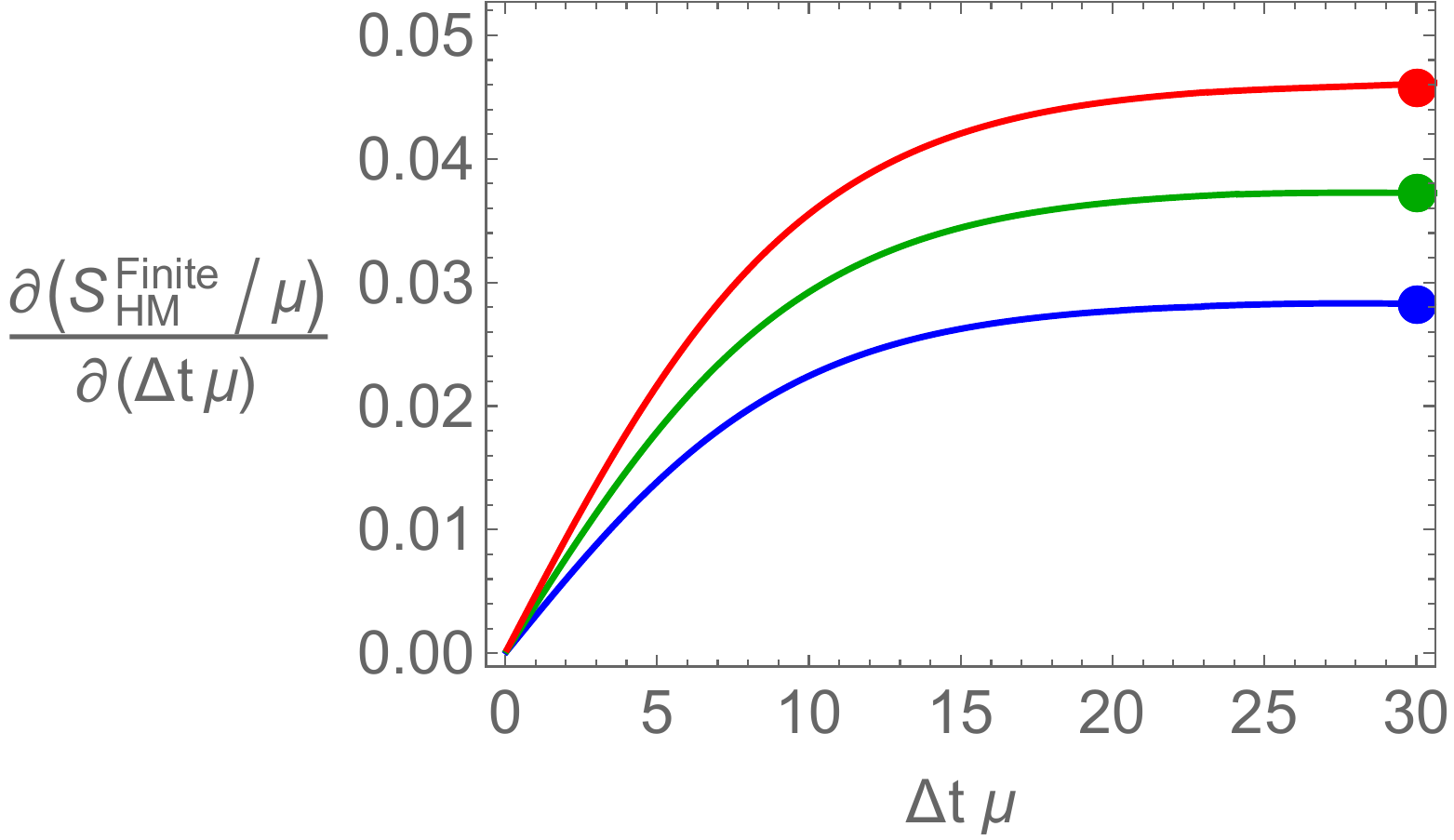} \label{}}
 \caption{The result from the time-dependent HM surface at $T/\mu=0.03$ with $c_{pb}=0, 1, 1.5$ (blue, green, red). The left figure shows the time evolution of $S^{\text{Finite}}_{\text{HM}}(t)$, while the right figure is its growth rate over time. In all figures, the solid lines are numerical results and dots are analytic results: \eqref{TRIVIAL1} (left), \eqref{LTAA} (right).}\label{HMTFIG1}
\end{figure*}

Let us first discuss the time-dependent HM surface, $S^{\text{Finite}}_{\text{HM}}(t)$ in \eqref{FORMU}.
Implementing the method in section \ref{sec32}, we find the time evolution of $S^{\text{Finite}}_{\text{HM}}(t)$ at given temperature. See Fig. \ref{HMTFIG1}.

In the left figure of Fig. \ref{HMTFIG1}, the numerical results (solid lines) show that $S^{\text{Finite}}_{\text{HM}}$ grows in time where its $\Delta t \rightarrow 0$ limit is consistent with the analytic results (dots) evaluated from \eqref{TRIVIAL1}. We also find that at given time, $c_{pb}$ enhances the value of $S^{\text{Finite}}_{\text{HM}}$ (e.g., from blue to red). 

On the other hand, in the opposite limit $\Delta t \rightarrow \infty$, one can find that $S^{\text{Finite}}_{\text{HM}}$ exhibits a monotonically (linearly) increasing behavior.
This linear behavior is more visible in the time derivative of entropy: see the right figure in Fig. \ref{HMTFIG1}.
We also check that our numerical results (solid lines) are consistent with the previously derived analytic expression (dots) of the late time analysis of HM surface, given in~\cite{Ling:2020laa}:
\begin{align}\label{LTAA}
\begin{split}
  \lim_{\Delta t \rightarrow \infty} \frac{\partial S_{\text{HM}}^{\text{Finite}}}{\partial \Delta t}  =  \frac{\sqrt{-f(z_M)}}{z^{d-1}_M} \,, 
\end{split}
\end{align}
where $z_M$ can be determined by solving
\begin{align}\label{}
\begin{split}
    (1-d) z_M^{-d}
    \sqrt{-f\left(z_M\right)}-\frac{z_M^{1-d
    } f'\left(z_M\right)}{2
    \sqrt{-f\left(z_M\right)}} = 0.
\end{split}
\end{align}

\begin{figure}[]
\centering
     {\includegraphics[width=8.5cm]{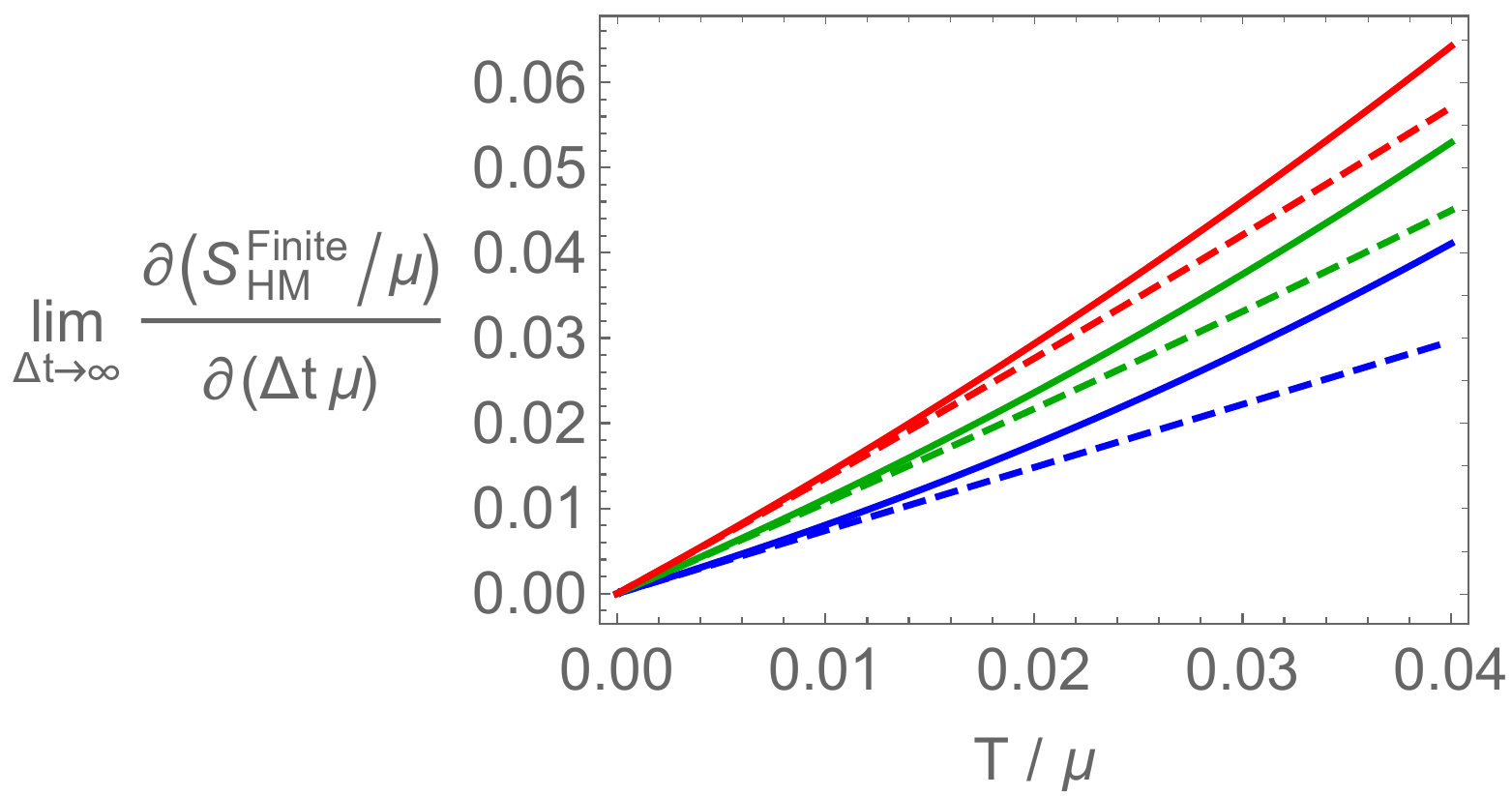} \label{}}
 \caption{The temperature dependence in $S^{\text{Finite}}_{\text{HM}}$ with $c_{pb}=0, 1, 1.5$ (blue, green, red). Solid lines are from \eqref{LTAA} and the dashed lines are its low temperature limit \eqref{CHFOR22}. The data at $T/\mu=0.03$ corresponds to the dots in the right figure of Fig. \ref{HMTFIG1}.}\label{TEMDEPFI}
\end{figure}

We also study the temperature dependence in $S^{\text{Finite}}_{\text{HM}}$: see Fig. \ref{TEMDEPFI}. 
We find that the entropy increases at higher temperatures or equivalently, it barely grows in the low-temperature limit.
In particular, we also find that the growth rate is linear in temperature in the near extremal limit:
\begin{align}\label{CHFOR22}
\begin{split}
\lim_{\Delta t \rightarrow \infty} \frac{\partial S_{\text{HM}}^{\text{Finite}}}{\partial \Delta t} \,=\, \frac{\pi}{3} \sqrt{\frac{1+c_{pb}^2 (T=0)}{2}} \,\frac{T}{\mu} \,+\, \mathcal{O}\left(\frac{T}{\mu}\right)^2  \,, 
\end{split}
\end{align}
where $c_{pb}^2 (T=0)$ is from \eqref{CHFOR}. Once the topological coefficient vanishes, it reduces to the one in \cite{Ling:2020laa}. Also it is believed that the origin of this linearity can be attributed to the infrared (IR) geometry~\cite{Ling:2020laa}.\footnote{One can find (e.g., see \cite{Jeong:2021zsv}) that the dyonic black hole still has AdS$_2\times R^2$ IR geometry as in the charged black hole. It would also be interesting to study how the IR parameters such as the hyperscaling violation exponent and the critical dynamical exponent change the temperature scaling in \eqref{CHFOR22}. See \cite{Jeong:2022jmp} for one of the recent studies of the entanglement entropy with IR parameters.}

As the entanglement between the black hole and the radiation is established through the exchange of Hawking modes before the Page time, the observed temperature dependence implies that a higher Hawking temperature corresponds to a greater rate of exchange.

One can easily find $\Delta S_{\text{HM}}(t)$ in \eqref{FORMU} using the data in the left figure of Fig. \ref{HMTFIG1}.
Also notice that such an entropy from the HM surface will keep growing due to the exchange of Hawking mode and exceed the maximum entropy of the black hole, which is in contrast with what the unitarity principle imposes as demonstrated in the introduction. The other candidate of the extremal surface, the island surface, will contribute to having the saturated maximum entropy and resolve this information paradox.

\subsection{Entanglement entropy from island surface}
Based on the result $S^{\text{Finite}}_{\text{HM}}(t=0)$ above, next we study the entanglement entropy from the island surface, $\Delta S_{\text{Is}}$ in \eqref{FORMU}, using the holographic formula \eqref{ISLAND1}.
Recall that as described in the previous section, $\Delta S_{\text{Is}}$ is time-independent, $\Delta S_{\text{Is}}(t) = \Delta S_{\text{Is}}(t=0)$, since the extremal surface is not associated to the stretch of space inside the horizon.

Furthermore, also recall that $\Delta S_{\text{Is}}$ may have an issue for the Page curve in doubly holographic theories: the positive sign of it, $\Delta S_{\text{Is}}>0$, is not guaranteed. Note that if its sign is negative (i.e., the dominant entropy at $t=0$ is from the island surface), one cannot have the Page curve since the entropy is already being saturated at $t=0$.
In order to resolve this issue, it is suggested \cite{Ling:2020laa} that one needs to consider the large value of $x_b$, i.e., moving the end point of the radiation region away from the Planck brane (e.g., see the right figure of Fig. \ref{SK223FIG}).

\begin{figure*}[]
\centering
     {\includegraphics[width=6.7cm]{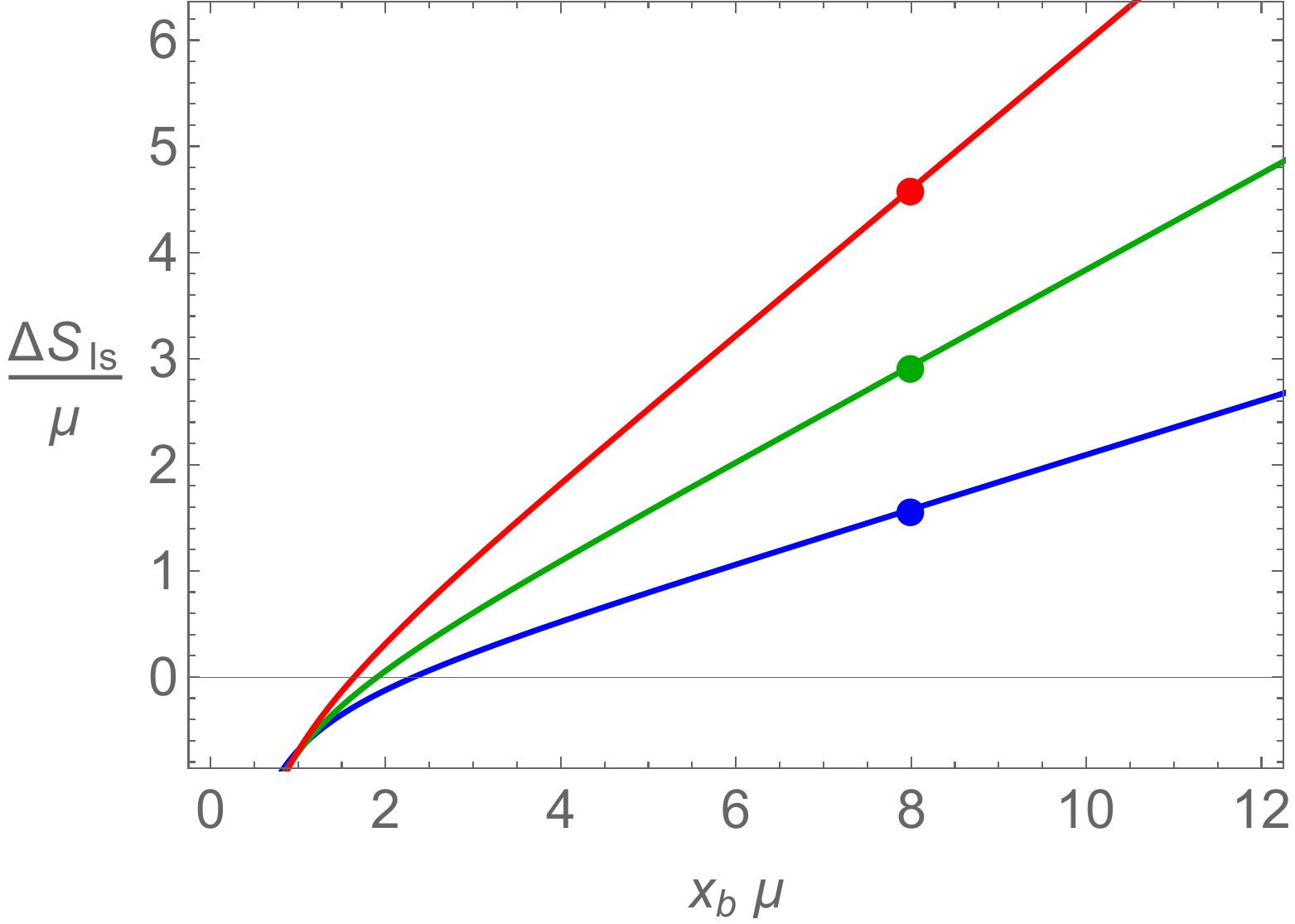} \label{}}
\qquad
     {\includegraphics[width=6.8cm]{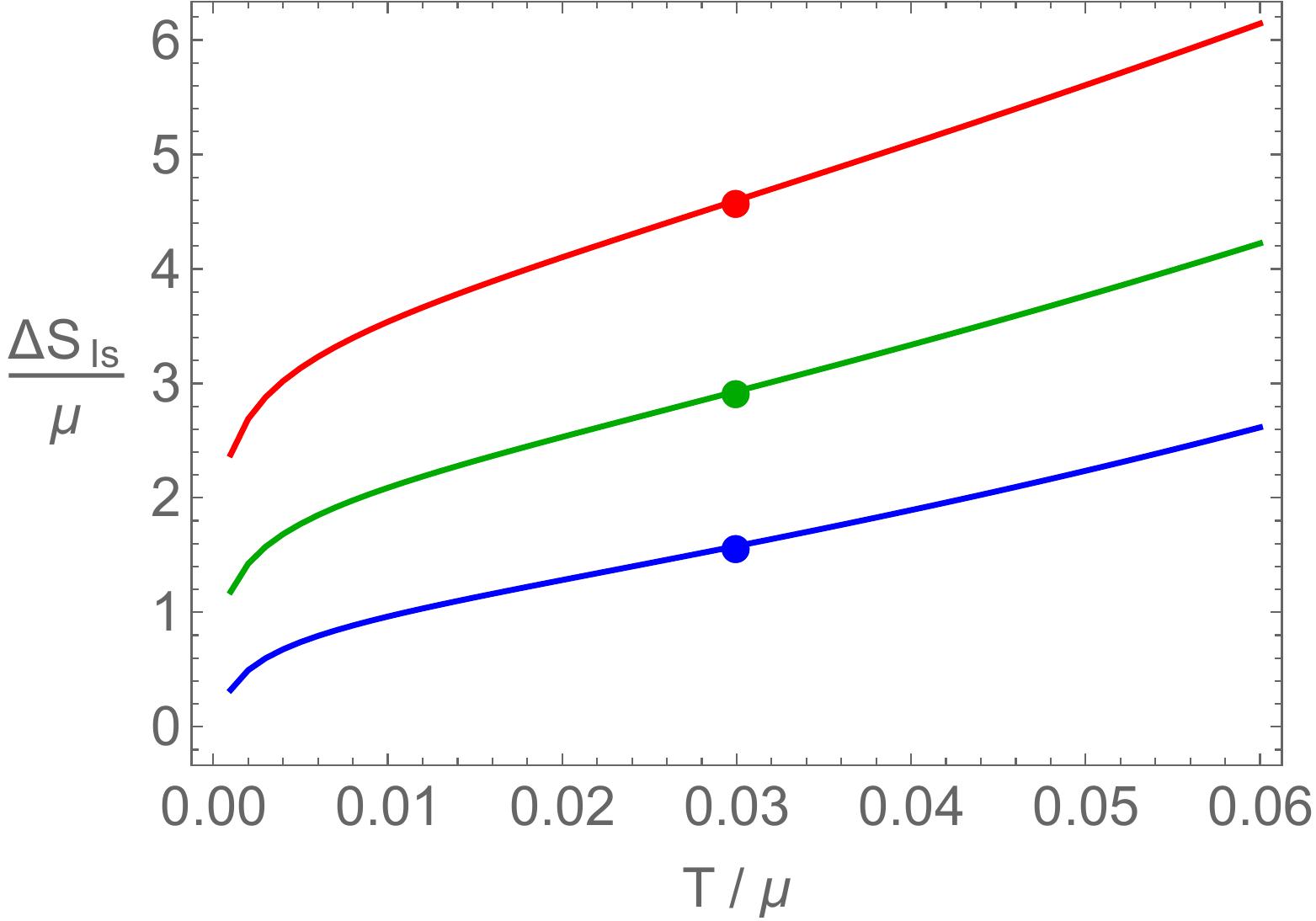} \label{}}
 \caption{The result from the island surface at $c_{pb}=0, 1, 1.5$ (blue, green, red). The left displays $\Delta S_{\text{Is}}$ at given $T/\mu = 0.03$ and the right shows the temperature dependence at given $x_b \mu = 8$. The dots are the same data in both figures.}\label{ISENTFIG}
\end{figure*}

In Fig. \ref{ISENTFIG}, we plot $\Delta S_{\text{Is}}$ for dyonic black holes.
In the left figure, we display the $x_b$ dependence at given $T$: one can check that $\Delta S_{\text{Is}}>0$ at larger values of $x_b$. We also present the $T$ dependence in the right figure, which is similar to the entropy from the HM surface: $T$ enhances the entropy.
In addition, we also observe that the role of $c_{pb}$ is also qualitatively the same: at given $x_b$ or $T$, the entropy is enhanced by a finite $c_{pb}$ (e.g., from blue to red).

\begin{figure}[]
\centering
     {\includegraphics[width=6.8cm]{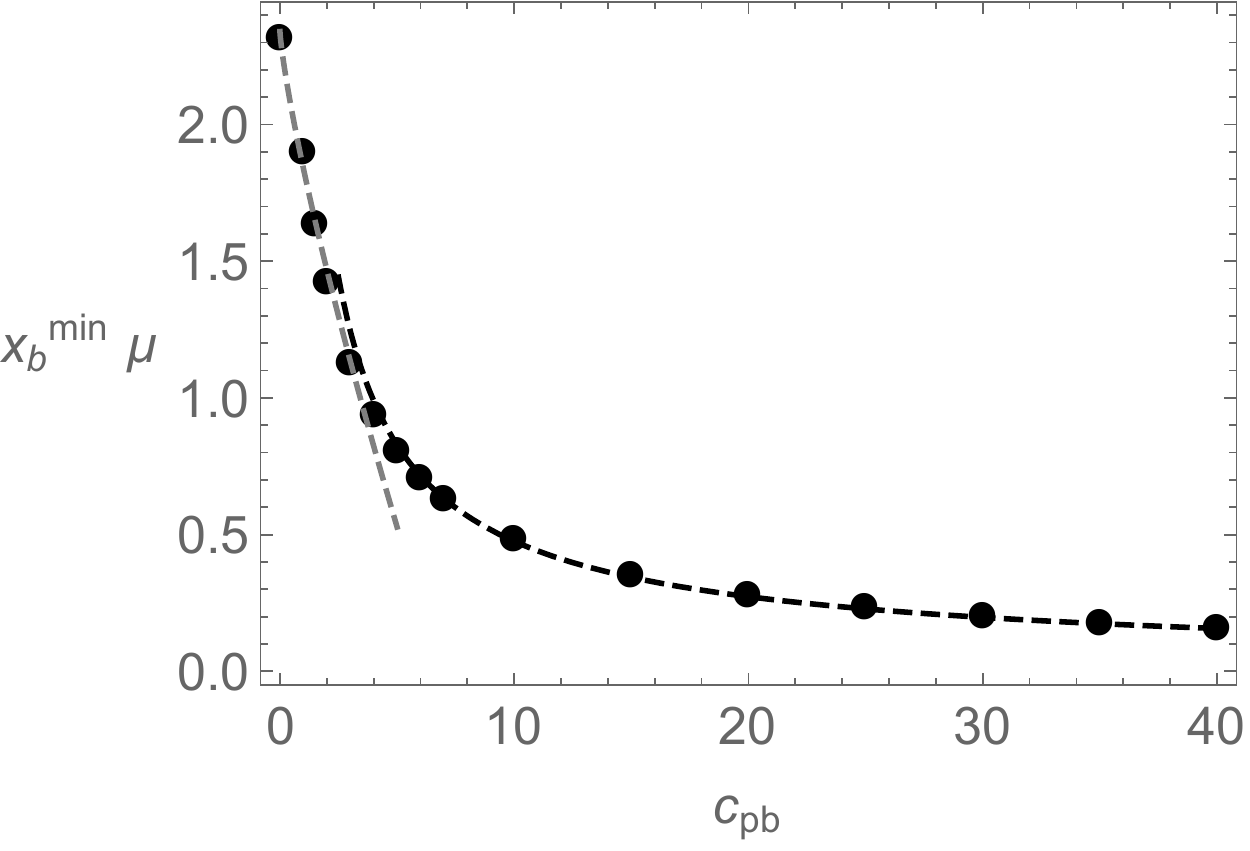} \label{}}
 \caption{$x_b^{\text{min}}$ vs. $c_{pb}$ at $T/\mu=0.03$. Dots are numerical results and dashed lines are fitting curves: $x_b^{\text{min}}\mu = 2.34-0.5 \,c_{pb}^{0.8}$ (gray), $3 \,c_{pb}^{-0.8}$ (black). }\label{CXBFIG2}
\end{figure}

It is also interesting that the coefficient $c_{pb}$ may also resolve the issue of the sign of $\Delta S_{\text{Is}}$ rather than taking a larger $x_b$.
In the left figure of Fig. \ref{ISENTFIG}, one can find that there is a minimum value of $x_b$ ($=:x_b^{\text{min}}$) at given $c_{pb}$, which is giving $\Delta S_{\text{Is}}=0$. For instance, $x_b^{\text{min}}\mu\approx2.34$ when $c_{pb}=0$ (blue). 
Such a minimum value depends on the value of $c_{pb}$, in particular, it is vanishing as we increase $c_{pb}$: see also Fig. \ref{CXBFIG2}.
This implies that one can have $\Delta S_{\text{Is}}\geq0$ in the all range of $x_b$ once we take the large enough $c_{pb}$.\footnote{{In order to avoid $\Delta S_{\text{Is}}<0$ (or the constant entropy belt), one may make use of a high tension brane or the large enough $c_{pb}$. Exploring the physical implications of $\Delta S_{\text{Is}}<0$ and the precise conditions necessary to attain a sufficiently large $c_{pb}$ presents an intriguing avenue for further investigation.}}\\

\subsection{Page curve of dyonic black holes}

\begin{figure*}[]
\centering
     {\includegraphics[width=6.8cm]{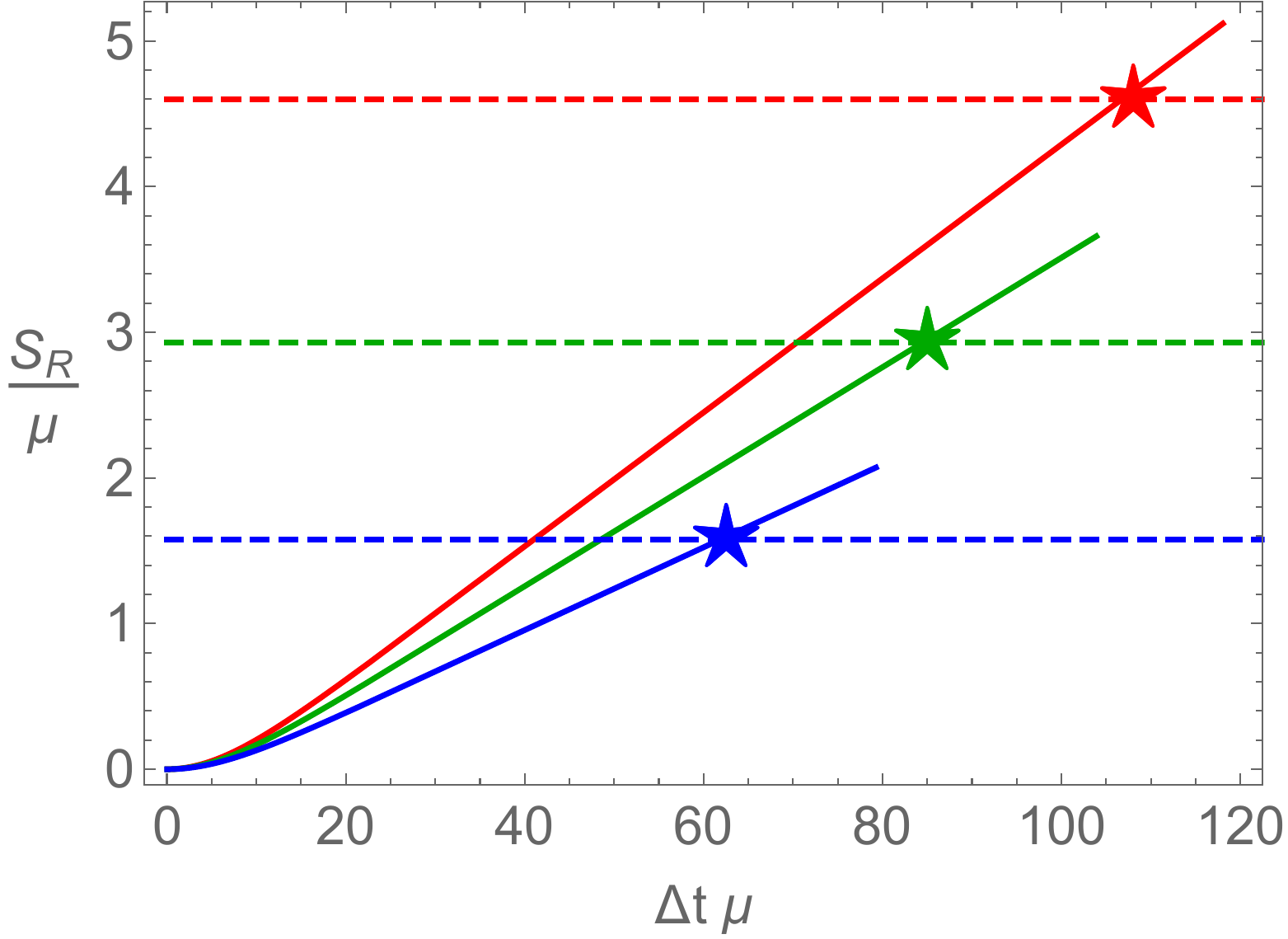} \label{}}
\qquad
     {\includegraphics[width=7.6cm]{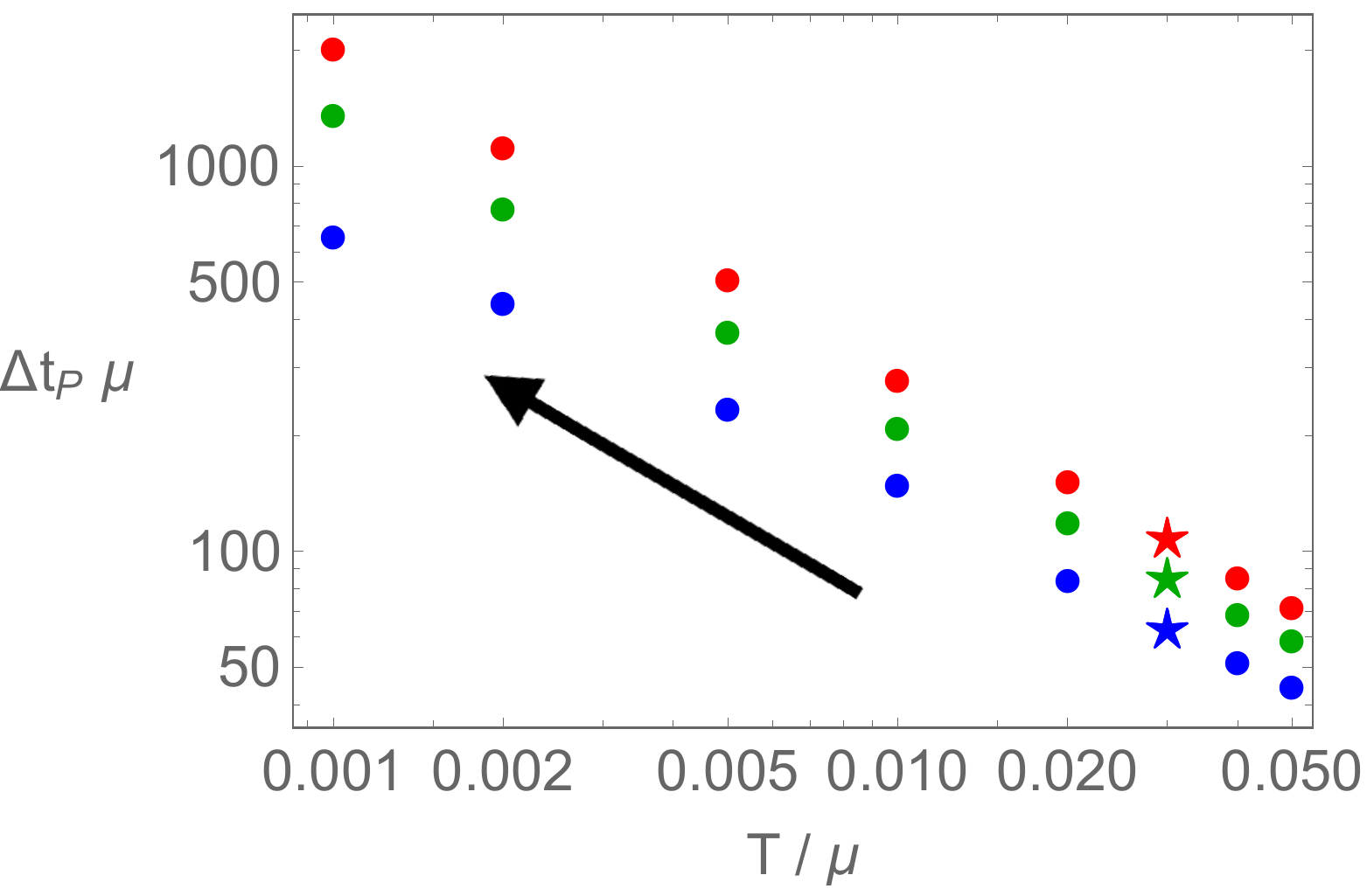} \label{}}
 \caption{The left figure displays the Page curve of dyonic black holes at $T/\mu=0.03$, $x_b\mu=8$ with $c_{pb}=0, 1, 1.5$ (blue, green, red). The solid line is $\Delta S_{\text{HM}}$, while the dashed line is $\Delta S_{\text{Is}}$. The {Page} time $t_P$ is depicted by the stars. The right figure shows the $T$ dependence in $t_P$. The stars indicate the same data in both figures.}\label{PCFIG1}
\end{figure*}

Finally, we discuss the Page curve of dyonic black holes using the evaluated entropy above: $\Delta S_{\text{HM}}$ and $\Delta S_{\text{Is}}$.
In the left figure of Fig. \ref{PCFIG1}, we display the Page curve \eqref{OVERALLPIC} where the entropy is described by the HM surface, $S_{\text{R}} = \Delta S_{\text{HM}}$ (solid lines), before the {Page} time (stars) and it is saturated by the island surface $S_{\text{R}} = \Delta S_{\text{Is}}$ (dashed lines) after the {Page} time.
Thus, the figure implies that the Page curve can be obtained even at finite $c_{pb}$.

In the right figure of Fig. \ref{PCFIG1}, we also show the $T$ dependence on the {Page} time $t_P$. We find that $t_P$ is enlarged in the low $T$ regime, which is consistent with the electrically charged black holes in the presence of a weak tension~\cite{Ling:2020laa}.\footnote{\label{ft15}However, recall that as we demonstrated in detail in section \ref{sec2la}, one cannot turn on a finite (even weak) tension on the brane for the purely electrically charged black holes \eqref{FDR}. See also footnote \ref{ft9}.} 
It is worth noting that this result, $t_P \approx 1/T$, is qualitatively in agreement with the {Page} time obtained from alternative approaches that do not utilize the doubly holographic method, for instance~\cite{Hashimoto:2020cas,Karananas:2020fwx,Wang:2021woy,Kim:2021gzd,Ahn:2021chg}.\\

\noindent \textbf{Entanglement density and the refined Page curve.}
Although the doubly holographic theories yield the Page curve exhibiting an initial growth in entropy, which subsequently saturates after the Page time, its effectiveness may be limited by the fact that the Page curve for an eternal black hole should saturate to twice the Bekenstein-Hawking entropy $S_{\text{BH}}$~\cite{Almheiri:2019psy,Ling:2020laa}\footnote{\label{ft16}It should be noted that the previous studies, for instance \cite{Almheiri:2019psy,Ling:2020laa}, did not report such saturation $S_{\text{R}} \approx 2 S_{\text{BH}}$ by the explicit calculations.
Furthermore, it can also be noted that in the doubly holographic setup, the $(d+1)$-dimensional horizon can be associated with the $d$-dimensional horizon at the brane~\cite{Almheiri:2019psy}. Consequently, the Bekenstein-Hawking entropy may also be applicable in this scenario.}
\begin{align} \label{SBHFOR}
\quad S_{\text{BH}} = \frac{L^{d-1}}{4 G_{N}} \frac{\sqrt{f_1(z_h)^{d-1}}}{z_{h}^{d-1}} \,,
\end{align}
where we use \eqref{GENMET}. It is also noteworthy that $S_{\text{R}}$ does not even have the same energy dimension with $S_{\text{BH}}$, for instance, in AdS$_4$, $S_{\text{R}}/\mu$ is dimensionless, while $S_{\text{BH}}/\mu^2$ is.

In this paper, inspired by the holographic entanglement density~\cite{Gushterov:2017vnr,Erdmenger:2017pfh,Giataganas:2021jbj,Jeong:2022zea}, we will show that taking a large $x_b$ limit can be useful not only to retain $\Delta S_{\text{Is}}\geq0$, but also to obtain $2 S_{\text{BH}}$ in the Page curve.

Let us first shortly review the holographic entanglement density below. For this purpose, it is convenient to rewrite the entanglement entropy at $t=0$, \eqref{ISLAND1} and \eqref{TRIVIAL1}, as follows:
\begin{align}\label{}
\begin{split}
S_{\text{Is}}   &\,=\,  \frac{2 L^{d-1} \Omega^{d-2}}{4 G_{N}}  \left[   \frac{1}{d-2} \frac{1}{{\epsilon}^{\,d-2}}   \,+\,  \frac{x_{b}\sqrt{f_1(z_b)^{d-1}}}{z_{b}^{d-1}}  \,+\, \frac{C_{\text{Is}}}{z_{b}^{d-2}}  \right]  \,, \\
S_{\text{HM}}   &\,=\,  \frac{2 L^{d-1} \Omega^{d-2}}{4 G_{N}}  \left[   \frac{1}{d-2} \frac{1}{{\epsilon}^{\,d-2}}   \,-\, \frac{1}{d-2}\frac{1}{z_h^{d-2}} \,+\, C_{\text{HM}}
  \right]  \,,
\end{split}
\end{align}
where $x_b$ is from \eqref{ISLAND3} and we defined
\begin{widetext}
\begin{align}\label{}
\begin{split}
C_{\text{Is}} &\,:=\, - \frac{1}{d-2}  \,+\, 
\int_{0}^{1} \frac{\dd u}{u^{d-1}}  \left( \sqrt{1-u^{2(d-1)}\frac{f_1(z_b)^{d-1}}{f_1(z_b u)^{d-1}}}\sqrt{f_2(z_b u) \, f_1(z_b u)^{d-2}}-1  \right)  \,, \\
C_{\text{HM}} &\,:=\, \int_{\epsilon}^{z_h} \frac{\dd u}{u^{d-1}}\left( \sqrt{f_2(u)\,f_1(u)^{d-2}} -1\right) \,, \\
\end{split}
\end{align}
\end{widetext}
Then, following \eqref{FORMU}, we can find the time-independent entanglement entropy responsible for the saturated Page curve after the {Page} time as 
\begin{align} \label{DENFOR11}
\begin{split}
\Delta S_{\text{Is}} \,&=\,S_{\text{Is}} \,-\, S_{\text{HM}} \\\,&=\,  \frac{2 L^{d-1} \Omega^{d-2}}{4 G_{N}}  \biggl[   \frac{x_{b}\sqrt{f_1(z_b)^{d-1}}}{z_{b}^{d-1}}  \,+\, \frac{C_{\text{Is}}}{z_{b}^{d-2}}  \\
& \qquad\qquad\qquad \,+\, \frac{1}{d-2}\frac{1}{z_h^{d-2}} -\, C_{\text{HM}} \biggr] \,.
\end{split}
\end{align}

\begin{figure*}[]
\centering
     {\includegraphics[width=6.8cm]{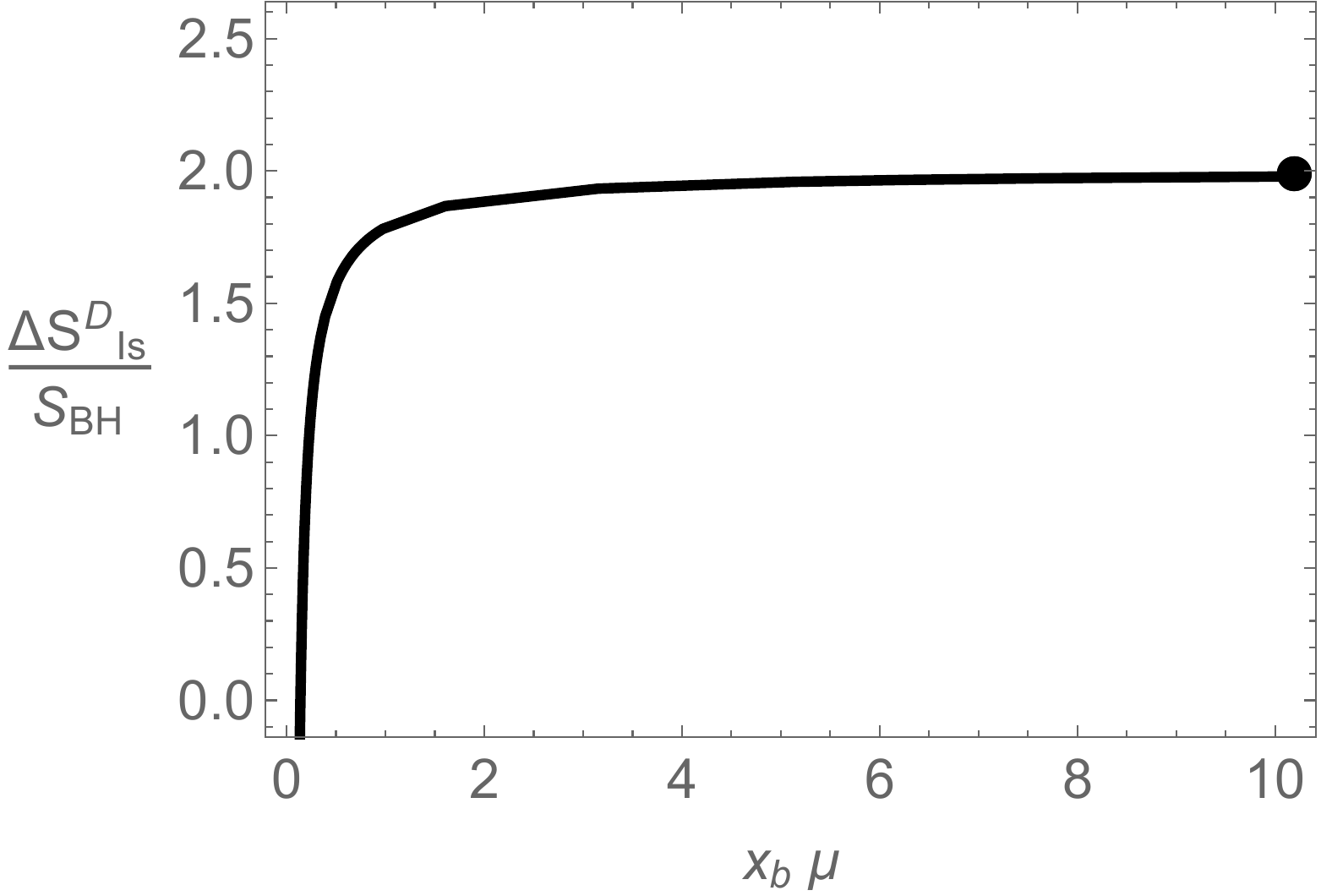} \label{}}
\qquad
     {\includegraphics[width=6.8cm]{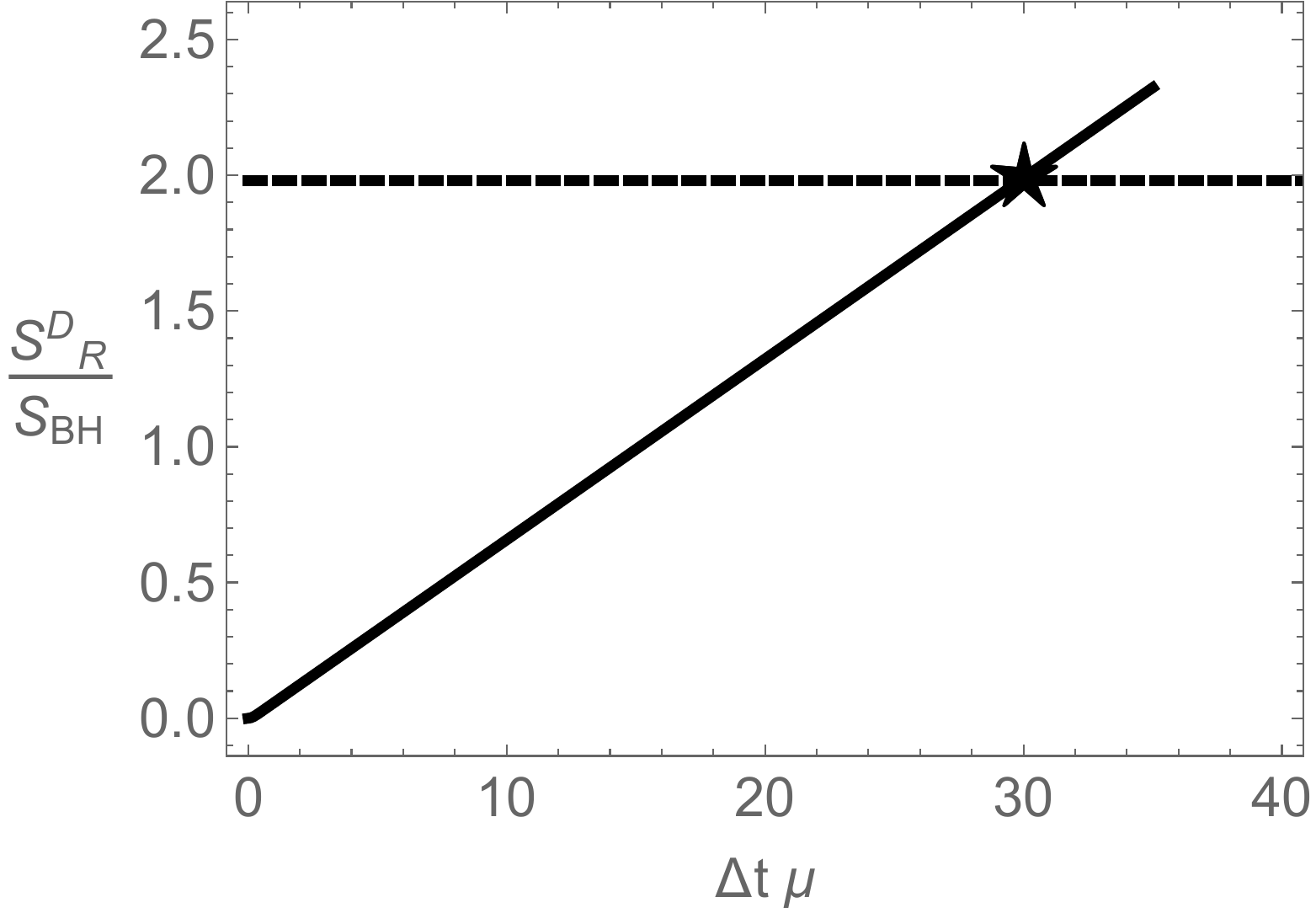} \label{}}
 \caption{Page curve and entanglement density at $c_{pb}=T/\mu=1$. In the left figure, $\Delta S^{D}_{\text{Is}} $ approaches $2 \, S_{\text{BH}}$ as we increase $x_b$, consistent with \eqref{DENFOR33}. In the right figure, we make a plot of the Page curve at  $x_b\mu=10$ where the star denotes the {Page} time.}\label{PRDFIG1}
\end{figure*}

The holographic entanglement density~\cite{Gushterov:2017vnr,Erdmenger:2017pfh,Giataganas:2021jbj,Jeong:2022zea} is defined by the holographic entanglement entropy divided by the volume of the boundary region $x_b \Omega^{d-2}$ as
\begin{align} \label{DCdc}
\Delta S^{D}_{\text{Is or HM}} \,:=\, \frac{\Delta S_{\text{Is or HM}}}{x_{b} \, \Omega^{d-2}} \,.
\end{align}
%
In this entanglement density context, \eqref{DENFOR11} is rewritten as 
\begin{align} \label{DENFOR22}
\begin{split}
\Delta S^{D}_{\text{Is}} &\,=\,  \frac{2L^{d-1}}{4 G_{N}}  \biggl[   \frac{\sqrt{f_1(z_b)^{d-1}}}{z_{b}^{d-1}}  \\
& \qquad\qquad \,+\,
\frac{1}{x_{b}} \left( \frac{C_{\text{Is}}}{z_{b}^{d-2}}  \,+\, \frac{1}{d-2}\frac{1}{z_h^{d-2}} \,-\, C_{\text{HM}} \right) \biggr] \,.
\end{split}
\end{align}
As we take the large $x_b$ limit, one can expect that the maximum value of $z_b$ approaches the horizon, i.e., $\lim\limits_{x_{b} \rightarrow \infty} z_{b} = z_{h}$~\cite{Gushterov:2017vnr,Erdmenger:2017pfh,Giataganas:2021jbj,Jeong:2022zea}.
Then, the leading contribution of \eqref{DENFOR22} is
\begin{align} \label{DENFOR33}
\begin{split}
\Delta S^{D}_{\text{Is}} \,\approx\,  \frac{2L^{d-1}}{4 G_{N}}     \frac{\sqrt{f_1(z_h)^{d-1}}}{z_{h}^{d-1}}  = 2 \, S_{\text{BH}} \,,
\end{split}
\end{align}
where $S_{\text{BH}}$ is from \eqref{SBHFOR}.
For instance, we display the representative actual data in the left figure of Fig. \ref{PRDFIG1}. 

The appearance of the Bekenstein-Hawking entropy in the large $x_b$ limit, \eqref{DENFOR33}, can be attributed to the volume law term in the standard RT surface: when $x_b$ is large, the subsystem becomes the entire system so that the minimal surface lies along the horizon~\cite{Hubeny:2012ry, Liu:2013una}. This implies that the ordinary RT surface (or island surface here) becomes the thermal entropy density $S_{\text{BH}}$ in \eqref{SBHFOR}. 
The factor ``2" of $2 S_{\text{BH}}$ is from the fact that our doubly holographic setup is for the thermofield double state, i.e., the ``two"-sided eternal black holes.
See also footnote \ref{ft16}.
For more detailed description of the volume law term (as well as the area law term associated with the area theorem~\cite{Ryu:2006ef,Myers:2012ed,Casini:2012ei,Casini:2016udt}), see \cite{Jeong:2022zea}.

{
One remark is in order. As demonstated in \cite{Chen:2020hmv}, the entanglement entropy from the island surface can be matched with the Bekenstein-Hawking entropy in the context of doubly holographic models: for instance, when the tension on the brane is large enough the extremal surface in the $d$-dimensional theory can be close to the horizon on the brane. Essentially, this scenario serves as the $d$-dimensional description for the Bekenstein-Hawking entropy at late times. 

Nevertheless, the primary focus of this section is to present a method for finding the Bekenstein-Hawking entropy even in the tensionless limit. The key inquiry under consideration is whether it is feasible to locate the extremal surface close to the horizon on the brane in the tensionless limit.

To achieve this, employing the concept of entanglement density, our strategy is based on two observations:
(I) in the limit of large $x_b$, the island surface approaches the $(d+1)$-dimensional horizon;
(II) the $(d+1)$-dimensional horizon intersects with the brane at the $d$-dimensional horizon~\cite{Almheiri:2019psy} (also refer to footnote \ref{ft16}).
These observations indicate that the island surface may be associated with the $d$-dimensional horizon when we consider a large value of $x_b$. Consequently, this can also imply a connection between our $S_{\text{BH}}$ in \eqref{DENFOR33} and the $d$-dimensional description of the Bekenstein-Hawking entropy on the brane.\footnote{Our argument may share similarities with the maximum tension case discussed in \cite{Chen:2020hmv}. In \cite{Chen:2020hmv}, the area of the bulk RT surface within the ($d+1$)-dimensional bulk, denoted as $A_{\text{RT}}/(4G_N)$, can be dominated by a local contribution. This local contribution corresponds to the area of the intersection between the RT surface and the brane, represented as $A_{\text{QES}}/(4G_{\text{eff}})$, which signifies the quantum extremal surface (QES) within the $d$-dimensional theory governed by the effective Newton's constant $G_{\text{eff}}$. Extremizing the area of the bulk RT surface drives the QES in close proximity to the horizon on the brane, enabling us to determine the Bekenstein-Hawking entropy within the $d$-dimensional theory, expressed as $A_{\text{horizon}}/(4G_{\text{eff}})$.}
}

Therefore, if we refine the entanglement entropy of the radiation \eqref{OVERALLPIC} using the entanglement density \eqref{DCdc} as
\begin{align}\label{}
S^{D}_{\text{R}} \,=\,
\begin{cases}
\Delta S^{D}_{\text{HM}}(t) \,,    \quad       (t < t_{P})     \\
\Delta S^{D}_{\text{Is}}(t)  \,,      \quad\,\,\,\,     (t \geqslant t_{P})    
\end{cases}
\end{align}
we find the expected behavior of the Page curve of the eternal black hole, the entanglement entropy is saturated to twice the Bekenstein-Hawking entropy after the {Page} time, in the context of doubly holographic theories: see also the right figure of Fig. \ref{PRDFIG1}.

%
\section{Conclusion}\label{sec5la}
We have studied the entanglement between the eternal black hole and the Hawking radiation in the context of the doubly holographic theories.  In particular, we consider the entanglement entropy of the radiation and aim to find the Page curve consistent with the unitarity principle. The main implication of the doubly holographic theories is that the ordinary RT/HRT prescription (using two types of extremal surfaces: Hartman-Maldacena surface and island surface) can yield a positive resolution of the information paradox through the appearance of the island.

The doubly holographic method for the Page curve is initially  implemented in the lower-dimensional black hole~\cite{Almheiri:2019hni} and subsequently extended to the higher-dimensional black holes: neutral black hole~\cite{Almheiri:2019psy} and charged black hole~\cite{Ling:2020laa}.
In this paper, following the previous literature, we study the Page curve of the dyonic black holes within doubly holographic theories.

We find that the extension to include a finite magnetic field would be a non-trivial task in that dyonic black hole in the doubly holographic setup requires the additional topological actions on the Planck brane~\cite{Fujita:2012fp,Melnikov:2012tb} where its effective topological coefficient is denoted as $c_{pb}$ in \eqref{SNEQ}.

Note that by virtue of a finite $c_{pb}$, the density has a relation with the magnetic field \eqref{RATD2}. In addition, analyzing such topological actions in detail, we also find that both the tension of the Planck brane and $c_{pb}$ should vanish for the purely electrically charged black holes. See also footnote \ref{ft9} and footnote \ref{ft15}.

Furthermore, we also develop a general (in metric and dimension) numerical method to compute the time-dependent Hartman-Maldacena surface, which produces numerical results in excellent agreement with analytic expressions.

Considering the tensionless brane but finite $c_{pb}$, we find that the doubly holographic theories can exhibit the Page curve consistent with the unitarity principles for the dyonic black holes: the entanglement entropy grows at early time and saturates after the {Page} time. The initial growth can be explained by the Hartman-Maldacena surface, while the saturation is attributed to the island surface.
As a byproduct, we also find that the aspect of the obtained {Page} time is consistent with the one derived from other approaches that do not employ the doubly holographic method in the literature.

Finally, using the holographic entanglement density~\cite{Gushterov:2017vnr,Erdmenger:2017pfh,Giataganas:2021jbj,Jeong:2022zea}, we also demonstrate that the saturated value of the entanglement entropy after the {Page} time can be comparable to twice the Bekenstein-Hawking entropy. To our knowledge, our work is the first doubly holographic study showing this twice the Bekenstein-Hawking entropy by explicit calculations.

Therefore, our analysis of dyonic black holes in the doubly holographic framework may provide another concrete example to support that the island paradigm would be a general solution to the information paradox for black holes in higher dimensions.

There can be a natural extension of our work. It may be desirable to investigate or develop a new method to include the effect of the tension on the Planck brane. For instance, see \cite{Fujita:2012fp,Melnikov:2012tb,Ling:2020laa} for some discussion on the finite tension brane.
In general, at a finite tension, the Planck brane is likely to give the back-reaction to the background geometry. In such a case, one may need to explore the entanglement entropy of the radiation beyond the scope of the way that we presented in this paper. 

It will also be interesting to study other quantum information quantities (such as the subregion complexity, reflected entropy) from dyonic black holes in the framework of doubly holographic theories and compare/{contrast} with the entanglement entropy given in this work. We leave this subject as future work and will address them in the near future.

%
\section*{Acknowledgments}
We would like to thank {D.S. Ageev, Daniel Arean, I.Ya. Aref'eva, Sang-Eon Bak, A.I. Belokon, Teng-Zhou Lai, Yi Ling, Yan Liu, Yuxuan Liu, Cheng Peng, V.V. Pushkarev, T.A. Rusalev, Zhuo-Yu Xian} for valuable discussions/correspondence and to {Yongjun Ahn} for earlier collaboration on related topics.
This work was supported by Project No. 12035016 supported by National Natural Science Foundation of China, the Strategic Priority Research Program of Chinese Academy of Sciences, Grant No. XDB28000000, Basic Science Research Program through the National Research Foundation of Korea (NRF) funded by the Ministry of Science, ICT $\&$ Future Planning (NRF-2021R1A2C1006791) and the AI-based GIST Research Scientist Project grant funded by the GIST in 2023.
This work was also supported by Creation of the Quantum Information Science R$\&$D Ecosystem (Grant No. 2022M3H3A106307411) through the National Research Foundation of Korea (NRF) funded by the Korean government (Ministry of Science and ICT).
H.-S Jeong acknowledges the support of the Spanish MINECO ``Centro de Excelencia Severo Ochoa'' Programme under Grant No. SEV-2012-0249. This work is supported through Grants No. CEX2020-001007-S and PID2021-123017NB-I00, funded by MCIN/AEI/10.13039/501100011033 and by ERDF A way of making Europe.\\

H.-S.J., K.-Y.K. and Y.-W.S. contributed equally to this paper.

%
\bibliographystyle{apsrev4-1}

\providecommand{\href}[2]{#2}\begingroup\raggedright\endgroup

\end{document}